\documentclass[prb,showpacs,twocolumn,superscriptaddress,amsmath,amssymb,floatfix]{revtex4-1}
\usepackage{graphicx}
\usepackage{bm}
\usepackage{bbm}
\usepackage{color}
\usepackage{amsmath}
\usepackage{amssymb}
\usepackage{diagbox}
\usepackage{color}
\usepackage{tabu} 
\usepackage{footnote}
\usepackage{epstopdf}
\usepackage{multirow} 
\usepackage{makecell} 
\usepackage{hyperref}
\usepackage{appendix}
\usepackage[applemac]{inputenc}

\begin{document}
\title{Odd-frequency superconducting pairing in Kitaev-based junctions}
\author{Athanasios Tsintzis}
\affiliation{Department of Physics and Astronomy, Uppsala University, Box 516, S-751 20 Uppsala, Sweden}
\affiliation{Division of Solid State Physics and NanoLund, Lund University, Box 118, S-221 00 Lund, Sweden}
\author{Annica M. Black-Schaffer}
\author{Jorge Cayao}
\affiliation{Department of Physics and Astronomy, Uppsala University, Box 516, S-751 20 Uppsala, Sweden}
\date{\today} 


\begin{abstract}
We investigate odd-frequency superconducting correlations in normal-superconductor (NS) and short superconductor-normal-superconductor (SNS) junctions with the S region described by the Kitaev model of spinless fermions in one dimension. 
We demonstrate that, in both the trivial and topological phases, Andreev reflection is responsible for the coexistence of even- and odd-frequency pair amplitudes in N and S close to the interfaces, while normal reflections additionally only contributes to  odd-frequency pairing in S. 
In the S region close to the NS interface we find that the odd-frequency pair amplitude exhibits large, but finite, values  in the topological phase at low frequencies due to the emergence of a Majorana zero mode at the interface which also spreads into the N region.
We also show that in S both the local density of states and local odd-frequency pairing can be characterized solely by  Andreev reflections deep in the topological phase. 
Moreover, in the topological phase of short SNS junctions, we find that both even- and odd-frequency amplitudes  capture the emergence of topological Andreev bound states. For a superconducting phase difference $0<\phi<\pi$ the odd-frequency magnitude exhibits a linear frequency ($\sim |\omega|$) dependence at low-frequencies,  while at $\phi=\pi$ it develops a resonance peak ($\sim 1/|\omega|$) due to the protected Majorana zero modes.
\end{abstract}

\maketitle

\section{Introduction}
\label{sect0}
The properties of a superconductor are to a very high degree given by the symmetries of the electron pair amplitude. Due to Fermi-Dirac statistics, the pair amplitude is necessarily antisymmetric under the total exchange of all quantum numbers. In the simplest situation, these degrees of freedom include time (or frequency), spin, and spatial coordinates. 
Conventionally, electron pairing occurs between electrons at equal times, leading to pairing with even-frequency, spin-singlet, and even parity (e.g.~the conventional superconducting state) or even-frequency, spin-triplet, and odd parity symmetries. However, in its more general form, pairing can occur at different times and also be odd in the relative time, or equivalently frequency, coordinate, enabling the possibility of odd-frequency superconducting pairing, such as odd-frequency, spin-singlet, and odd-parity or odd-frequency, spin-triplet, and even parity symmetries.

Odd-frequency pairing was initially considered as an intrinsic phenomenon in superfluid Helium
\cite{bere74} and later investigated  for superconductivity.\cite{PhysRevLett.66.1533,PhysRevB.45.13125,PhysRevB.46.8393,PhysRevB.48.7445,PhysRevB.52.1271,0953-8984-9-2-002,PhysRevB.60.3485}  This exotic pairing can also arise as an intrinsic effect in multiband superconductors\cite{PhysRevB.88.104514,PhysRevB.92.094517,PhysRevB.92.224508,PhysRevLett.119.087001,PhysRevB.97.064505}  or it can be induced  in superconducting hybrids or junctions,\cite{PhysRevLett.86.4096,PhysRevLett.98.037003,PhysRevLett.99.037005,Eschrig2007,PhysRevB.75.134510,PhysRevB.76.054522,Nagaosa12,PhysRevB.86.144506,PhysRevB.87.104513,
PhysRevB.87.220506,PhysRevB.92.014508,PhysRevB.92.121404,PhysRevB.92.205424,PhysRevB.92.100507,PhysRevB.92.134512,Ebisu16, PhysRevB.95.184518,PhysRevB.98.161408,
PhysRevB.95.224502,bo2016,PhysRevB.96.155426,PhysRevB.96.174509,PhysRevB.97.075408,PhysRevB.97.134523,Triola18,PhysRevLett.120.037701,PhysRevB.98.075425,PhysRevB.99.184501} as well as under time-dependent fields.\cite{PhysRevB.94.094518,triola17} 

A large part of the relevance of odd-frequency pairing comes from allowing for highly unconventional superconducting correlations. For example, experiments have demonstrated that odd-frequency pairing enables a long-range proximity effect in ferromagnet-superconductor junctions.\cite{longrangeExp,bernardo15}  In this case, the ferromagnet induces a spin-singlet to triplet conversion while still keeping the robust $s$-wave spatial parity, thus allowing superconductivity to survive and propagate in the ferromagnet. Similarly, in non-magnetic junctions, odd-frequency pairing is generated due to  spatial parity breaking at the interface, which allows the conversion from $s$-wave to $p$-wave symmetry, while conserving the spin-singlet structure.\cite{PhysRevB.76.054522} For a recent review on progress on odd-frequency superconductivity, see Ref.\,\onlinecite{Balatsky2017}.

Moreover, zero bias peaks have been reported in conductance measurements on many unconventional superconductors.\cite{PhysRevB.23.5788,doi:10.1143/PTP.76.1237,PhysRevLett.72.1526,PhysRevLett.74.3451,PhysRevB.53.9371,PhysRevB.56.892,0034-4885-63-10-202} In some cases, such peaks correspond to the emergence of Majorana zero modes (MZMs) that signal a topological superconducting phase.\cite{PhysRevLett.89.077002,kitaev,PhysRevLett.105.077001,PhysRevLett.105.177002,PhysRevB.83.224511,Nagaosa12, Aguadoreview17,LutchynReview08} The MZM is a particle equal to its antiparticle and has attracted a great deal of attention due to its potential application in fault tolerant quantum computation.\cite{kitaev,RevModPhys.80.1083,Sarma:16,PhysRevB.98.161408,PhysRevLett.121.267002} 
It has been shown that the frequency behavior of the pair amplitude produced by MZMs is highly unusual in the sense that it always exhibits an odd-frequency dependence.\cite{PhysRevB.87.104513} This result is supported by further studies  based on a  quasiclassical approach in for example junctions with diffusive N regions,\cite{PhysRevLett.98.037003,PhysRevB.76.054522,Takagi18}  junctions with magnetic fields,\cite{PhysRevB.87.104513} and  systems deep in the topological phase with solely MZMs.\cite{PhysRevB.92.014513,PhysRevB.95.174516, lutchyn16, PhysRevB.96.155426,tamura18} 

The fact that MZMs automatically give rise to odd-frequency pairing has raised the notion that these two concepts are intimately intertwined. On the other hand, MZMs only appear at boundaries of topological systems, and at boundaries there can also be odd-frequency pairing appearing simply due to spatial parity breaking. It is therefore natural to ask the question  what is the relationship between MZMs and the associated  topological phase, and odd-frequency pairing?.
A detailed answer to this question is clearly important towards forming a more complete understanding of both odd-frequency pairing and MZMs. 

The simplest platform in which to investigate MZMs is the Kitaev model,\cite{kitaev} which describes a one-dimensional $p$-wave superconducting wire with spinless or spin-polarized fermions. This model exhibits a topological superconducting phase that is characterized by the emergence of MZMs in finite systems, one at each end of the wire, with their wavefunctions exponentially decaying into the bulk of the superconductor.\cite{kitaev} Despite its simplicity, the Kitaev model holds experimental relevance as its physical properties can be engineered by combining semiconducting nanowires with Rashba spin-orbit coupling and proximity-induced conventional $s$-wave superconductivity, where an external magnetic field drives the system into its topological phase.\cite{PhysRevLett.105.077001,PhysRevLett.105.177002} This extension of the Kitaev model has led to great experimental activity with remarkable results during the past few years.\cite{chang15, Higginbotham, Krogstrup15, zhang16, Albrecht16, Deng16, Nichele17, Suominen17, Chene1701476, Marcus2018nonlocality,zhang18} Junctions based on the Kitaev model, therefore, represents both the simplest and the most natural platform for analyzing MZMs and their relation to odd-frequency pairing.

In this work we study NS and short SNS Kitaev-based junctions by employing a fully analytical quantum mechanical  Green's function approach to investigate the presence of odd-frequency pairing and its relation to MZMs. 
On one hand, NS junctions allow for the simplest study of interface scattering processes, including Andreev reflections, and thus allow for establishing their detailed relation to odd-frequency pairing and the emergence of a single MZM at the interface. 
On the other hand, short SNS junctions host a single pair of Andreev bound states (ABSs) in the energy gap at finite phase difference $\phi$. These ABSs  develop a protected zero energy crossing at $\phi=\pi$ in the topological phase which reflects the formation of a MZM pair, thus offering a clean way to study  MZMs and odd-frequency pairing without the influence of any additional energy levels.

In general, we demonstrate that even- and odd-frequency pair amplitudes in the N and S regions coexist in both the trivial and topological phases due to Andreev reflections, which act as spatial parity mixers at interfaces. 
Normal reflections, however, solely contribute to odd-frequency pairing in the S regions, in contrast to junctions with conventional $s$-wave superconductors.\cite{PhysRevB.92.205424,PhysRevB.98.075425}
For NS junctions in the topological phase, odd-frequency pairing in S  is maximized around zero frequency due to the emergence of a MZM at the interface. However, it does not exhibit the $\sim1/|\omega|$ dependence associated with an isolated MZM.\cite{lutchyn16,Takagi18,tamura18}
Moreover, we also find that deep in the topological phase in S, Andreev reflection offers a simultaneous characterization of  LDOS,  conductance, and  local odd-frequency pairing at the interface.  Large values of LDOS or conductance in the low frequency topological regime therefore signify dominant odd-frequency correlations in S.

In short SNS junctions we show that both odd- and even-frequency amplitudes reveal the emergence of the topological ABSs at the junction. Interestingly, we  obtain for $0<\phi<\pi$ an odd-frequency magnitude with a linear frequency ($\approx|\omega|$) dependence at low-frequencies.  At $\phi=\pi$, however, the magnitude develops a resonance peak $\sim 1/|\omega|$, revealing the MZM pair,  which emerges due to the protected zero-frequency crossing at $\phi=\pi$ in this topological junction.  The observation of a zero-frequency resonance peak in the LDOS at $\phi=\pi$ at the junction thus clearly indicates the presence of large  odd-frequency correlations. This also applies to the $4\pi$-fractional Josephson effect as it occurs due to  the protected zero frequency crossing at $\phi=\pi$.\cite{PhysRevB.53.9371,kitaev,Kwon2004}

We thus conclude that odd-frequency pairing is not restricted to either the topological phase and its MZMs, but appears very generally also in the absence of MZMs. The appearance of MZMs does enhance the odd-frequency pairing, but notably does not always appear as a sharp $1/|\omega|$ peak, instead giving a much more subtle contribution at NS interfaces.

The remainder of this article is organized as follows. In Sec.\,\ref{sect1} we present the model and outline the employed method. In Sec.\,\ref{sec2} we present our results for NS junctions in the trivial and topological phases. We then, in the same section, study the pair amplitudes in SNS junctions in the topological phase. Finally, we present our conclusions in Sec.\,\ref{concl}. For completeness, we provide all the details on the derivation of the analytical calculations reported in this work in Appendices \ref{App1}-\ref{AppSNS}.

\section{Model}
\label{sect1}
\begin{figure}[!t]
\centering
\includegraphics[width=.49\textwidth]{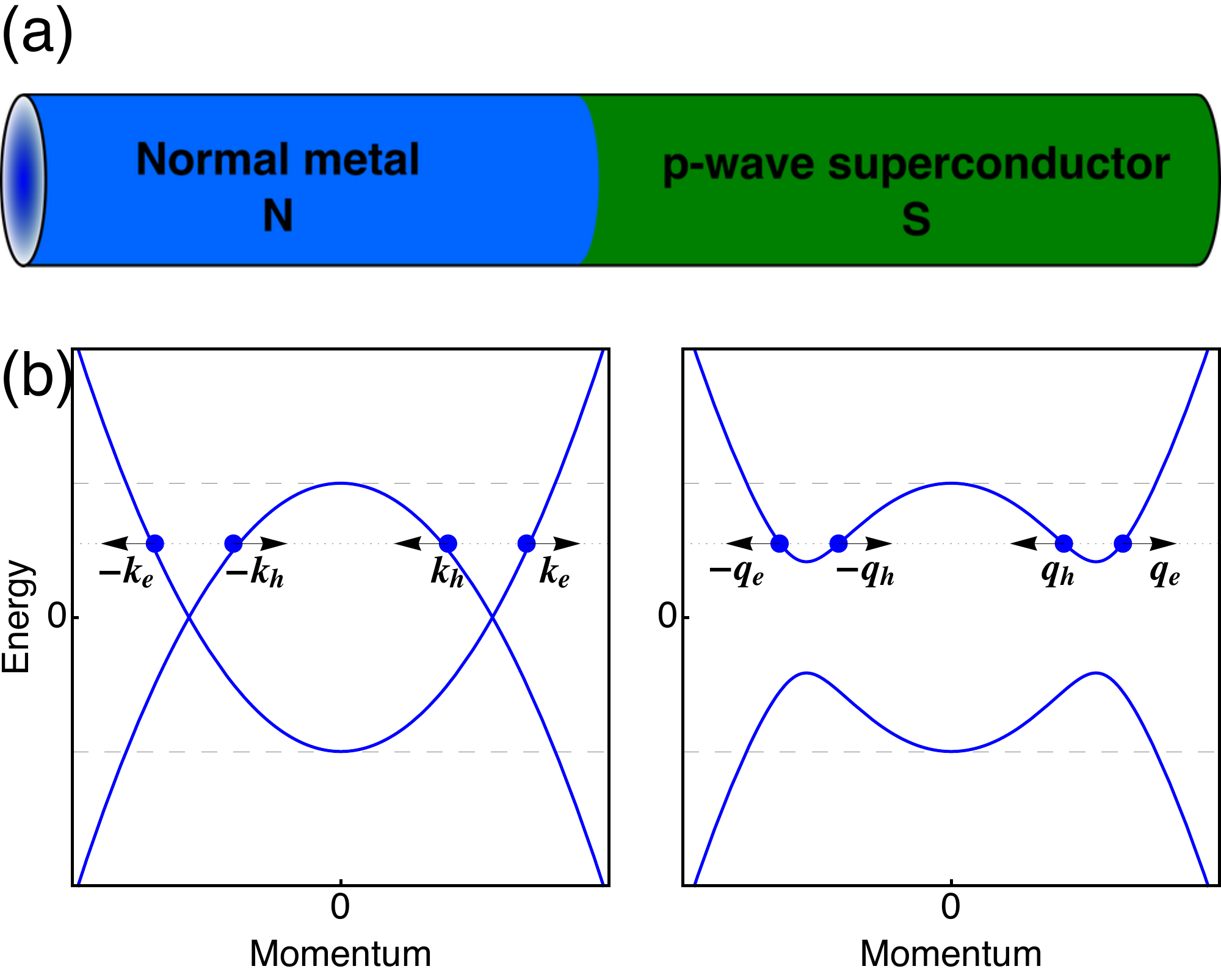} 
\caption{(Color online.) (a) Normal metal (N) in contact with a 1D $p$-wave superconductor (S) described by the Kitaev model. (b) Energy versus momentum dispersion at positive chemical potential (marked by dashed line) in the normal (left) and superconducting (right) regimes.   The left (right) pointing arrows represent left (right) moving electrons/holes with wavevectors $k_{e(h)}$ in N and $q_{e(h)}$ in S.}
\label{fig1}
\end{figure}

The purpose of this work is to investigate the induced superconducting pair correlations in one-dimensional (1D) NS and short SNS junctions described by the 1D Kitaev model.
As an example, a NS junction is schematically shown in Fig.\,\ref{fig1}(a), with the continuum Bogoliubov-de Gennes (BdG) Hamiltonian given by
\begin{equation} 
\label{eq1}
 H_{BdG}=
\begin{pmatrix}
H_{0} & \Delta_{p}p  \\
\Delta_{p}^* p  & - H_{0}
\end{pmatrix}.
\end{equation}
Here $H_{0}=\frac{p^{2}}{2m} -\mu_{i}+V\delta(x)$ is the kinetic term set by the effective mass $m$ and momentum $p=(-i\hbar\partial_{x})$, 
with $\mu_{i}$ being the chemical potential in the $i=\text{N(S)}$ normal (superconductor) region measuring the filling of the band, $V$  the strength of the interface barrier potential which controls the interface transparency, and $\Delta_{p}(x)=\theta(x)\Delta$ the $p$-wave pairing potential\footnote{Spatial variations of $\Delta(x)$ at the NS interfaces\cite{PhysRevB.97.155425} do not
alter the main conclusions of this work.} 
intrinsic to the Kitaev model with zero superconducting phase assumed for simplicity. Note that the N region is here spin-polarized or spinless, a property shared with the Kitaev model and also expected for a strong magnetic field applied along the wire needed to reach the topological phase in the S region. In what follows we use $\mu_{\rm S}=\mu$ for simplicity of notation. The short SNS junction  has a similar setup but with an additional S region with a superconducting phase difference $\phi$ between the two superconductors, and with the length of the normal part vanishing $L_N \rightarrow 0$.

Diagonalizing the Hamiltonian in Eq.\,(\ref{eq1}) in the normal state, $\Delta=0$,  gives the eigenvalues  
plotted in Fig.\,\ref{fig1}(b) with wavevectors 
\begin{equation}
\label{kN}
k_{e,h}=k_{\mu_{\rm N}}\sqrt{1\pm \frac{\omega}{\mu_{\rm N}}}\,,
\end{equation}
where $k_{\mu_{\rm N}}=\sqrt{2m\mu_{\rm N}/\hbar^{2}}$. These wavevectors characterize the right and left moving electrons (holes) indicated by filled blue circles with horizontal arrows in Fig.\,\ref{fig1}(b).  For $\mu_{\rm N}>0$ the normal bands correspond to two parabolas that cross at $\pm k_{\mu_{\rm N}}$ and describe a  metallic N region, while for $\mu_{\rm N}<0$ the N region describes an insulator. 

On the other hand, a finite $\Delta$ opens a gap in the spectrum. The Kitaev model is particularly interesting because it realizes a topological phase with a MZM at each end of the wire.\cite{kitaev} It has been shown that at $\mu=0$ the spectrum of the $p$-wave superconductor becomes gapless and signals the topological transition point between the trivial ($\mu<0$) and topological ($\mu>0$) phases.
The gap of the spectrum is $E_{1}=\sqrt{[2m\Delta^{2}/\hbar^{2}][\mu-m\Delta^{2}/(2\hbar^{2})]}$  for $\mu>m\Delta^{2}$ and $\mu$ for $0<\mu<m\Delta^{2}$, while in the trivial phase, $\mu<0$, the gap is $|\mu|$. The right panel of Fig.\,\ref{fig1}(b) shows the gapped spectrum in the topological phase, where it opens at the Fermi points $\pm k_{\mu_{\rm N}}$. The  wavevectors at finite $\Delta$  are given by
\begin{equation}
\label{wavevector}
q_{e,h}=\sqrt{\frac{2m}{\hbar^{2}}}\sqrt{\mu-\frac{m\Delta^{2}}{\hbar^{2}}\pm \sqrt{\omega^{2}-E_{1}^{2}}}\,.
\end{equation}
These wavevectors characterize the right and left moving quasielectrons (quasiholes) indicated by filled blue circles with horizontal arrows in right panel of Fig.\,\ref{fig1}(b).
Moreover, $q_{e(h)}$ can become complex valued depending of the values of $\omega$, $\mu$ and $\Delta$, as discussed when necessary. Here, the superconducting coherence length is given by $\xi=(\hbar v_{\mu})/(k_{\mu}\Delta)$, where $v_{\mu}=k_{\mu}/m$ and $k_{\mu}=\sqrt{2m\mu/\hbar^{2}}$, which reduces to $\xi=1/(m\Delta)$ for $\hbar=1$.

\section{Pair amplitude analysis}
\label{sec2}
In order to capture odd-frequency pairing we need to analyze the superconducting pair amplitudes, which can be calculated from the anomalous electron-hole Green's functions components. 
To this end, we first construct the retarded Green's functions $G^{r}(x,x',\omega)$ 
 with outgoing boundary conditions in each region derived from the scattering processes at the interface.\cite{PhysRev.175.559} 
 We here follow the approach developed in Refs.\,\onlinecite{PhysRevB.96.155426,PhysRevB.98.075425}, but for a comprehensive treatment, we provide the details in Appendix \ref{App1}. Note that this method is particularly suitable in this study because it both gives analytical access to the pair amplitudes and provides the scattering processes at the interfaces which hold experimental relevance in conductance measurements. Since the full derivations of the Green's functions $G^{r}$ are straightforward but tedious we give them in Appendices \ref{AppNS} and \ref{AppSNS}, for the NS and SNS systems, respectively.
 
With spin not being explicitly present in the system,  $G^{r}(x,x',\omega)$  is a $2\times2$ matrix in electron-hole space, whose off-diagonal element $G_{eh}^{r}(x,x',\omega)$  directly gives the pair amplitudes
\begin{equation}
\label{pairingF}
G_{eh}^{r}(x,x',\omega)=f^{r}(x,x',\omega)\,.
\end{equation}
Note that the spin-symmetry of the pair amplitude corresponds to the spin configuration in the Kitaev model, where the spinless consideration has to be understood in terms of a single spin species or a spin-polarized texture.
Therefore, the spin-symmetry of Eq.\,(\ref{pairingF}) automatically corresponds to the equal spin-triplet component ($\uparrow\uparrow$ or $\downarrow\downarrow$). 
The even and odd-frequency components of the pair amplitude are obtained from
\begin{equation}
\label{EOF}
\begin{split}
f^{r,{\rm E}}(x,x',\omega)&=\frac{f^{r}(x,x',\omega)+f^{a}(x,x',-\omega)}{2}\,,\\
f^{r,{\rm O}}(x,x',\omega)&=\frac{f^{r}(x,x',\omega)-f^{a}(x,x',-\omega)}{2}\,,
\end{split}
\end{equation}
where $f^{a}$ corresponds to the advanced pair amplitude and  is found using Eq.\,(\ref{pairingF}) from the relation $G^{a}(x,x',\omega)=[G^{r}(x',x,\omega)]^{\dagger}$. Here we have used that when considering negative frequencies ($\omega\rightarrow-\omega$) the advanced functions must be used instead of the retarded function.\cite{PhysRevB.96.155426} Note also that Eq.\,(\ref{EOF}) is valid for $\omega>0$, so if plotting results for $\omega<0$ we have to instead use the advanced counterparts. This is a standard procedure when working with retarded and advanced Green's functions, as for example described in Ref.\,\onlinecite{Balatsky2017}.
Throughout this work we not only study the pair amplitude $f^{r}$, but also  the pair magnitude which is defined as $|f^{r}|=\sqrt{(f^{r})(f^{r})^{*}}$.

\subsection{NS junctions}
\label{sec2A}
We start by considering NS junctions, as in Fig.\,\ref{fig1}(a), with the interface located at $x=0$, where the N ($x<0$) and S ($x>0$) regions are semi-infinite. 
First, we discuss the normal region and later the superconducting region. For the anomalous electron-hole  Green's function in N we obtain
\begin{equation}
\label{GehN_NS}
G^{r,{\rm N}}_{eh}(x,x',\omega) = \frac{i \eta }{2k_e}   r_{eh} \, {\rm e}^{-i(k_e x - k_h x')}\,,
\end{equation}
where $\eta=2m/\hbar^{2}$, $k_{e,h}$ are the electron and hole wavevectors defined in Eq.\,(\ref{kN}), and $r_{eh}$ is the Andreev reflection coefficient of a right moving electron into a left moving hole, which depends on the system parameters, see App.\,\ref{AppNS} for the detailed expression. 
Thus, the Andreev reflection, through $ r_{eh}$, is the unique scattering process that determines the anomalous Green's function in N, a behavior similar to NS junctions with conventional $s$-wave superconductors.\cite{PhysRevB.98.075425} The Andreev reflection is accompanied by an exponential term, with exponent $k_e x - k_h x'$, which mixes electron and hole wavevectors at different spatial coordinates. This exponential term, therefore, acts as a spatial parity mixer that enables the coexistence of even- and odd-frequency amplitudes at the interface. In order to visualize this, we decompose Eq.\,(\ref{GehN_NS}) into its even- and odd-frequency components in the large chemical potential limit, $\mu_{\rm N}\gg \omega$. In this regime, the wavevectors in Eqs.\,(\ref{kN}) are simply given by $k_{e,h}=k_{\mu_{\rm N}}+\kappa^{\rm N}_{\omega}$ with $\kappa^{\rm N}_{\omega}={k_{\mu_{\rm N}} \omega/2 \mu_N}$ and we obtain 
\begin{equation}
\label{pairingN_NS}
\begin{split}
f^{r,{\rm E}} (x,x',\omega) =&  \frac{\eta}{2k_e} r_{eh}  {\rm sin}[k_{\mu_{\rm N}}(x-x')]\, {\rm e}^{-i\kappa^{\rm N}_{\omega}(x+x')}\,,\\
f^{r,{\rm O}} (x,x',\omega) =& \frac{i\eta }{2k_e} r_{eh}  {\rm cos}[k_{\mu_{\rm N}}(x-x')]\, {\rm e}^{-i\kappa^{\rm N}_{\omega}(x+x')}\,. 
\end{split}
\end{equation}
We thus find that the coexistence of both symmetry classes is determined by Andreev reflection for $x\neq x'$.

 \begin{figure}[!t]
\centering
\includegraphics[width=.49\textwidth]{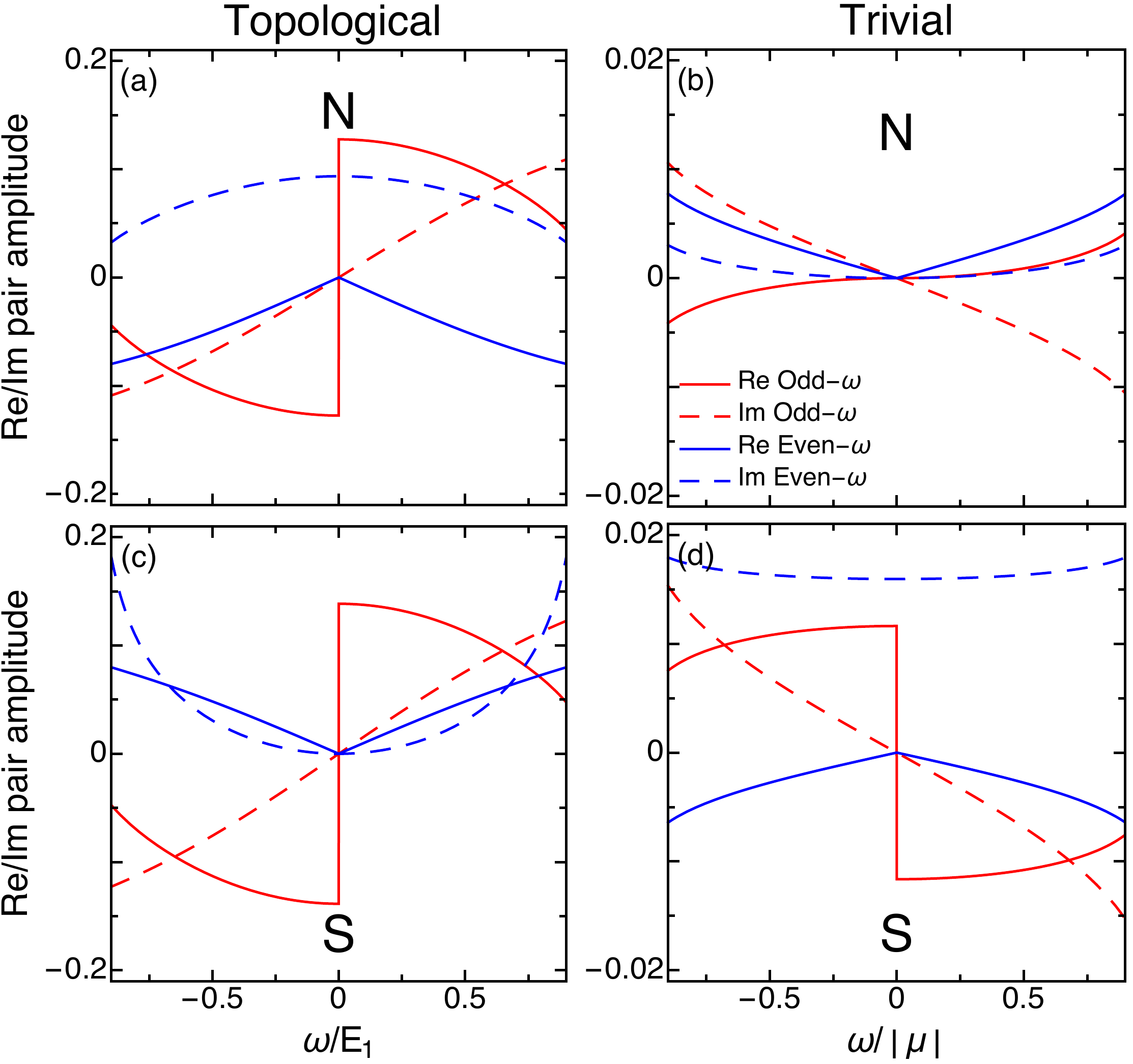} 
\caption{(Color online.) Real and imaginary parts of the interface pair amplitudes as a function of $\omega$ for a NS junction at $x'=0$ and for the N region at $x=-0.05\xi$ (a, b) and the S region at $x=0.05\xi$ (c, d). We choose $m=0.5$, $\Delta = 0.5$ and $V=0$. For the topological regime (a, c) $\mu = 4$ and for the trivial regime (b, d) $\mu = -1$. Note the different scales on the x-axis because of the normalization to the gap which is different in the topological and trivial regimes.}
\label{fig2}
\end{figure}

The even- and odd-frequency dependence of the pair amplitudes are plotted in Fig.2 (a,b) as a function of $\omega$ in both the topological and trivial phases (panels (c,d) for the S region ($x>0$) will be discussed in subsequent paragraphs). Note that for $\omega<0$ we plot the advanced pair amplitudes.
In the topological phase the imaginary part of the odd-frequency component develops a sharp sawtooth profile at $\omega=0$, as also found in topological insulators.\cite{PhysRevB.96.155426} In the trivial phase both amplitudes are smaller than in the topological phase, but still clearly finite, and smoothly evolve across $\omega=0$.
The odd- and even-frequency nature of the amplitudes in Eqs.\,(\ref{pairingN_NS}) is also consistent with their spatial dependence. The even-frequency amplitude is proportional to ${\rm sin}[k_{\mu_{\rm N}}(x-x')]$ and is therefore odd under the exchange of spatial coordinates $x$ and $x'$. The odd-frequency amplitude, on the other hand, is proportional to  ${\rm cos}[k_{\mu_{\rm N}}(x-x')]$,  which implies an even dependence under the exchange of $x$ and $x'$, in agreement with Fermi-Dirac statistics for spin-triplet pairs.

 \begin{figure}[!t]
\centering
\includegraphics[width=.49\textwidth]{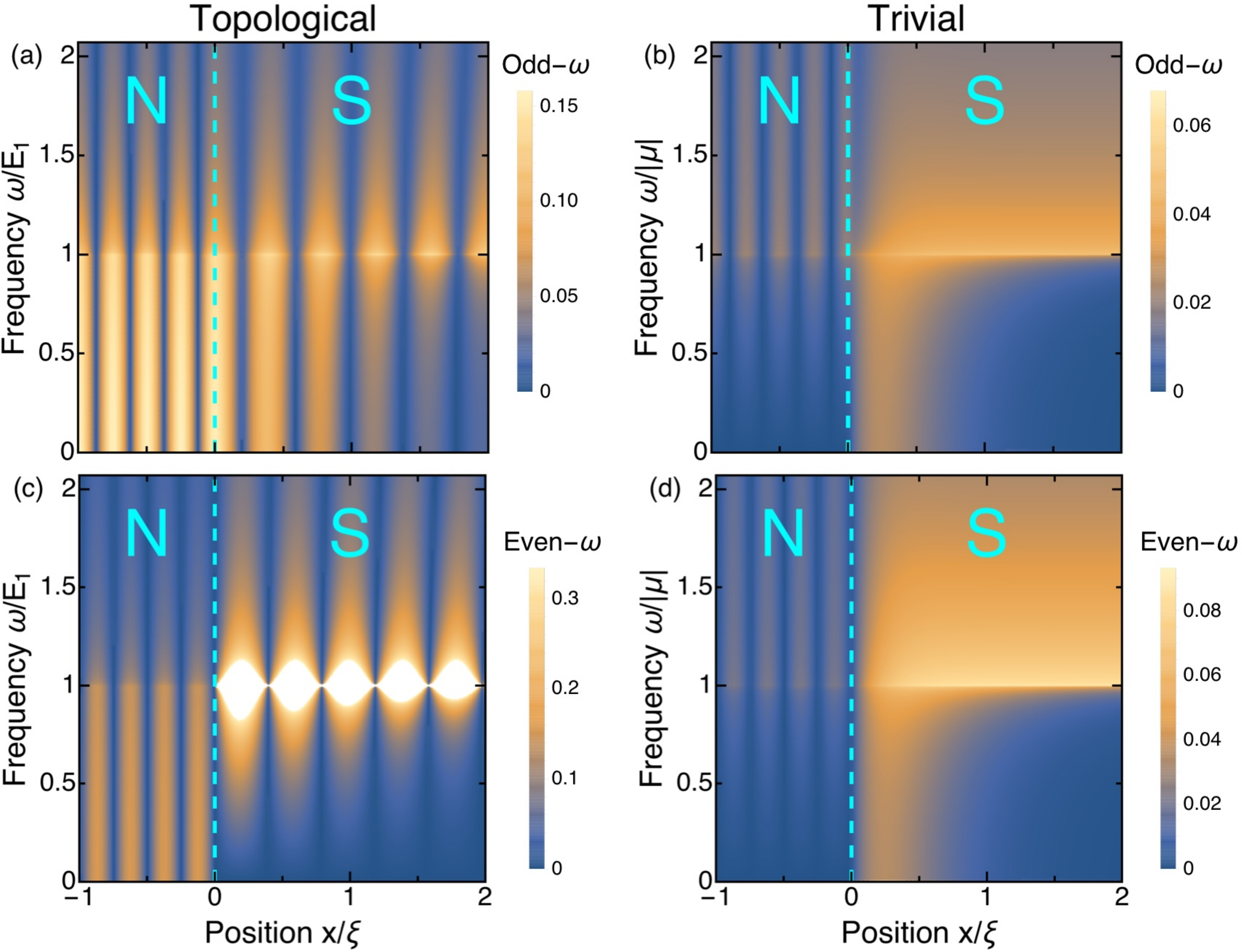} 
\caption{(Color online.) Odd- (a, b) and even-frequency (c, d) pair magnitudes as a function of $\omega$ and $x$ for a NS junction in the topological (a, c) and the trivial (b, d) regimes.  Dashed lines mark the NS interface. In the S region we plot only the interface components. Parameters same as in Fig. \ref{fig2}.}
\label{fig3}
\end{figure}

\begin{figure}[!t]
\centering
\includegraphics[width=.49\textwidth]{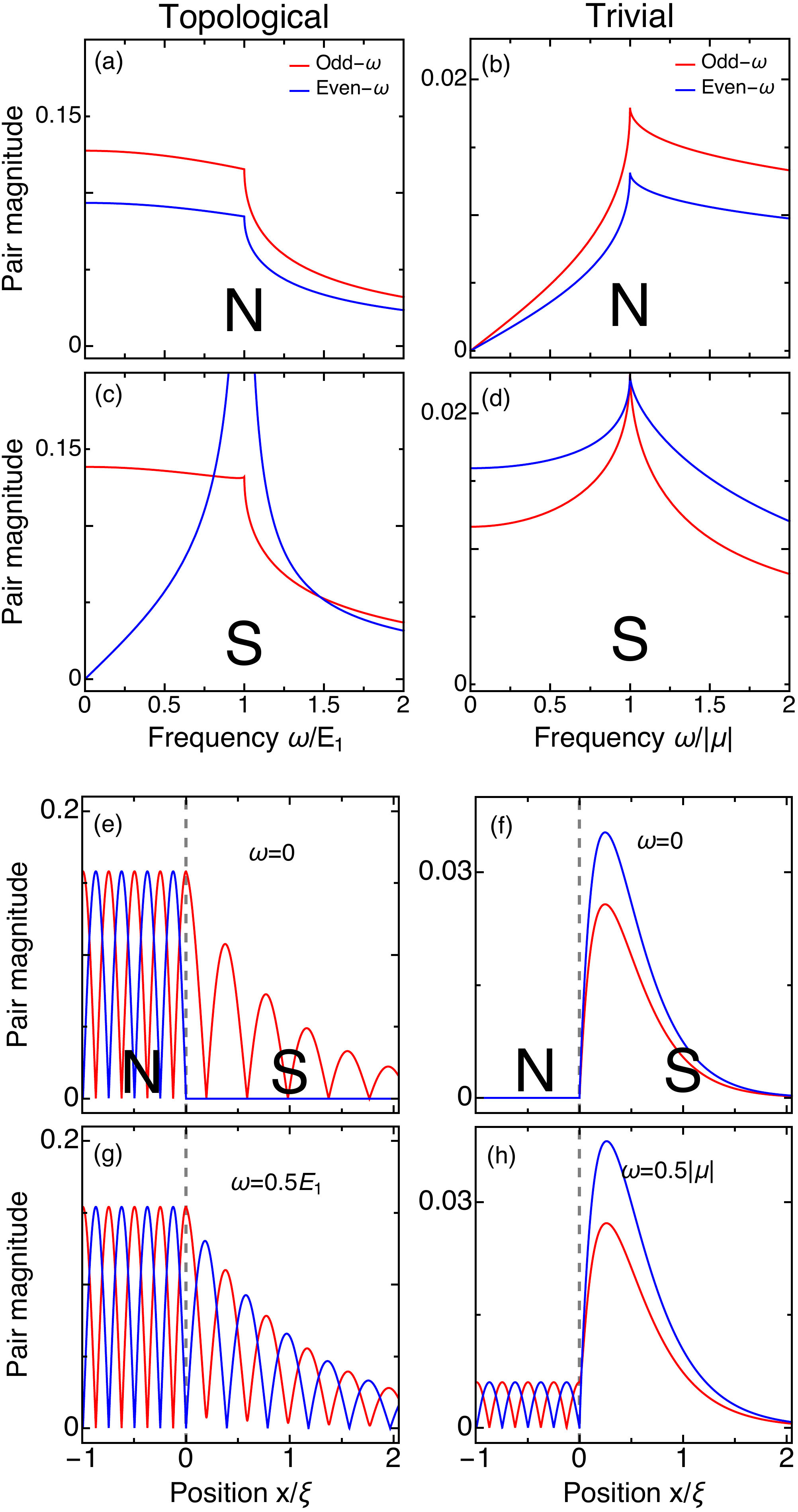} 
\caption{(Color online.) Frequency (a-d) and spatial (e-h) dependence of the odd- and even-frequency pair magnitudes for a NS junction in the topological and  trivial  regimes. For the S region we only plot the interface components. In panels (a-d)  we take $x'=0$, while $x=\mp0.05\xi$ for N(S). In panels (e-h) $x'=0$.
Vertical dashed lines in (e-h) mark the SN interface. Parameters same as in Fig. \ref{fig2}.}
\label{fig4}
\end{figure}

After confirming the odd- and even-frequency nature of the pair amplitudes in N, we plot in Fig.\,\ref{fig3} both the frequency and spatial dependence of their magnitudes. 
For a better visualization, we also extract in Fig.\,\ref{fig4}(a,b) the frequency dependent magnitudes for the topological and trivial regimes at fixed positions.  At $\omega\approx0$ in the topological phase both pair magnitudes are large with an approximate constant value for frequencies within the gap $E_{1}$, but decays fast for $\omega>E_{1}$, as  seen in Fig.\,\ref{fig4}(a). The finite value of the pair magnitudes in the topological phase is determined by the finite value of the Andreev reflection coefficient $r_{eh}$, as seen in Eqs.\,(\ref{pairingN_NS}). 
See also Fig.\,\ref{fig4}(e,g) for the oscillatory position dependence at fixed frequencies.
We have verified that by increasing the barrier strength $V$, the values of the pair amplitudes do not change at $\omega\approx0$, which stems from the large Andreev reflection due to the emergence of a MZM at the interface in the topological regime. 
At finite frequency, however, Andreev reflection gets reduced and therefore  both pair magnitudes acquire lower values. 

In the trivial phase both magnitudes are completely suppressed around zero frequency as seen in Figs.\,\ref{fig3}(b,d) and \ref{fig4}(b,f). This, again, results from the vanishing  Andreev reflection coefficient $r_{eh}$ in the trivial regime  at $\omega\approx0$ in N. As frequency increases, both pair magnitudes increase almost linearly with the odd-frequency exhibiting a faster raise as the frequency approaches the gap  $\mu$ in this regime.
Large values of the barrier strength tend to reduce the Andreev reflection coefficient $r_{eh}$, and therefore also reduce the pair amplitudes, albeit their overall behavior is preserved. 

In the superconducting region we obtain  for the anomalous electron-hole Green's function 
\begin{equation}
\label{GS_NS}
\begin{split}
&G^{r,{\rm S}}_{eh}(x,x',\omega)=\\
&= s  \Big[ {\rm e}^{-i|x-x'|q_{h}} - {\rm e}^{i|x-x'|q_{e}} \Big]
 {\rm sgn}(x-x') \\& 
+ s \Big[  r_{q_{h}q_{h}}  {\rm e}^{-i(x+x')q_{h}} - r_{q_{e}q_{e}}  {\rm e}^{i(x+x')q_{e}} \Big] \\
&+\frac{sV_{q_{h}} r_{q_{e}q_{h}}}{V_{q_{e}}U_{q_{e}}}  \Big[ U_{q_{h}} {\rm e}^{ i(q_{e} x'-q_{h}x)} -  U_{q_{e}} {\rm e}^{ i(q_{e}x - q_{h} x')}\Big]\,,
\end{split}
\end{equation}
where $s = [i \eta (1+U_{q_{e}}^2)U_{q_{h}}]/[2q_{h}(U_{q_{e}}^2 - U_{q_{h}}^2)]$, $U_{I}$, $V_{i}$ correspond to the coherence factors and the coefficients $r_{q_{h}q_{h}(q_{e}q_{e})}$ and $r_{q_{e}q_{h}}$ are  normal hole-to-hole (electron-to-electron) and Andreev reflection coefficients, respectively.  Their explicit expressions are given in Appendix \ref{AppNS}, but we discuss their overall behavior here. The anomalous Green's function, given by Eq.\,(\ref{GS_NS}), is composed of terms which allow us to determine their origin. In fact, the first term (first line)  does not exhibit any space dependence locally at $x=x'$, and therefore we associate it with the bulk of S. The second and third terms  are proportional to the normal and Andreev reflection coefficients, respectively, which originate from the interface. This observation allows us to distinguish different contributions\cite{PhysRevB.92.205424,PhysRevB.92.100507,PhysRevB.96.155426,PhysRevB.98.075425} to $G^{r}_{eh}$, which can be divided into terms coming from the bulk (first line) and from the interface (second and third lines) of S:
$f^{r}(x,x',\omega)= f^{r}_{\rm B}(x,x',\omega) + f^{r}_{\rm I}(x,x',\omega)$ with $f^{r}_{\rm I}(x,x',\omega)=f^{r}_{\rm NR}(x,x',\omega) + f^{r}_{\rm AR}(x,x',\omega)$, where NR (AR) denotes  normal (Andreev) reflection.

The bulk term in Eq.\,(\ref{GS_NS})   is odd under the exchange of spatial coordinates $(x,x')$, which corresponds to even-frequency symmetry, reflecting the nature of the parent $p$-wave superconductor in the Kitaev model.  On the other hand, the interface contributions from normal reflections are even under the exchange of spatial coordinates, and, therefore, are expected to contribute only to the odd-frequency component. Interestingly, the third term, which arises due to Andreev reflection, mixes spatial coordinates with electron and hole wavevectors, through $(-q_{h}x + q_{e} x')$ in the exponents, similarly to the behavior of the Andreev reflection in the N region.  
Thus, despite being  junctions of $p$-wave superconductors, the role of Andreev reflections, in the generation of even- and odd-frequency correlations, still behaves as in junctions with $s$-wave superconductors.\cite{PhysRevB.92.205424,PhysRevB.96.155426,PhysRevB.98.075425}
We further note that the interface components are particularly valuable because they can be characterized by means of  normal and Andreev reflections. These scattering processes hold experimental relevance as they can be directly obtained in conductance measurements.\cite{chang15,doi:10.1063/1.4971394,gulonder,Gazibegovic17,zhang18} In the following we thus focus on the interface terms, and using Eqs.\,(\ref{EOF}) we find the even- and odd-frequency interface pair amplitudes,
\begin{equation}
\label{soddeven}
\begin{split}
 f^{r,{\rm E}}_{\rm I} (x,x',\omega)&=  K_{\rm AR}^{+}\Big[ {\rm e}^{ i( q_{e} x'-q_{h}x)} -  {\rm e}^{ i(q_{e}x - q_{h} x')}\Big]\,,\\
f^{r,{\rm O}}_{\rm I} (x,x',\omega) &= K_{\rm AR}^{-}  \Big[ {\rm e}^{ i( q_{e} x'-q_{h}x)} +  {\rm e}^{ i(q_{e}x - q_{h} x')}\Big]\\
&+  \Big[  K_{\rm NR}^{h}  {\rm e}^{-i(x+x')q_{h}} -  K_{\rm NR}^{e} {\rm e}^{i(x+x')q_{e}} \Big]\,, 
\end{split}
\end{equation}
where $K_{\rm AR}^{\pm}=\pm\frac{s}{2}  \frac{U_{q_{e}} \pm U_{q_{h}}}{U_{q_{e}}}  \frac{V_{q_{h}}}{V_{q_{e}}}   r_{q_{e}q_{h}} $, $K_{\rm NR}^{e(h)}=s\,r_{q_{e}q_{e}(q_{h}q_{h})}$.
Observe that the Andreev reflection results in a coexistence of even- and odd-frequency components through coefficients $K_{\rm AR}^{\pm}$, while only the odd-frequency amplitude also has a contribution from the normal reflection. This is in stark contrast to  junctions with $s$-wave superconductors, where odd-frequency in S is solely proportional to the Andreev coefficient.\cite{PhysRevB.92.205424,PhysRevB.96.155426,PhysRevB.98.075425} 
Before going further, we verify the frequency dependence of both amplitudes and  plot their real and imaginary parts in Fig.\,\ref{fig2}(c,d) for $x>0$ in S. Observe that both  amplitudes exhibit the expected even- and odd-frequency dependence in the trivial and topological phases. Notice that in both trivial and topological phases, the real component of odd-frequency exhibits a sawtooth profile at $\omega\approx0$, while the other components have a smooth frequency dependence.

In order to gain additional insight into the behavior of the interface pair amplitudes given by Eqs.\,(\ref{soddeven}), we plot in Figs.\,\ref{fig3}-\ref{fig4} their magnitude in S as a function of both frequency and space. At $\omega\approx0$ in the topological phase in the S region the odd-frequency magnitude is maximum, while the even-frequency component completely vanishes. 
This can be understood by writing  Eqs.\,(\ref{soddeven}) in the topological regime in the large chemical potential limit, such that $q_{e,h}=k_{\mu}\pm i\kappa$ with $\kappa=|E_{1}|(k_{\mu}/\mu)$. Then Eqs.\,(\ref{soddeven}) read
\begin{equation}
\label{pairS}
\begin{split}
 f^{r,{\rm E}}_{\rm I} (x,x',\omega)&=  -2iK_{\rm AR}^{+}{\rm sin}[k_{\mu}(x-x')]{\rm e}^{-\kappa(x+x')}\,,\\
f^{r,{\rm O}}_{\rm I} (x,x',\omega) &= 2K_{\rm AR}^{-}  {\rm cos}[k_{\mu}(x-x')]{\rm e}^{-\kappa(x+x')}\\
&-2i K_{\rm NR}^{h}{\rm sin}[k_{\mu}(x+x')]{\rm e}^{-\kappa(x+x')}\,, 
\end{split}
\end{equation}
where we have used that in this regime $K_{\rm NR}^{h}=K_{\rm NR}^{e}$. The coefficients in Eqs.\,(\ref{pairS}) can be further simplified. In fact, in this topological regime at $\omega\approx0$ and $\mu>0$, the coherence factors read $U_{q_{e},q_{h}}=\pm i$, $V_{q_{e}}=V_{q_{h}}$ and thus by plugging them into the expressions for  $s$ and $K_{\rm AR}^{\pm}$ in Eq.\,(\ref{pairS}), we obtain vanishing interface even-frequency component $f_{\rm I}^{r,{\rm E}}=0$ but a finite value for the interface odd-frequency component $f_{\rm I}^{r,{\rm O}}$, which for $\omega\approx0$ has its maximum value at $x=0$. 
 When moving away from the interface, the odd-frequency component exhibits an exponential and oscillatory decay into S, while the even-frequency component remains at zero, as illustrated in Fig.\,\ref{fig4}(e). The decay length is  determined by the inverse of $\kappa$, which in this regime is $1/\kappa= [1/E_{1}|(k_{\mu}/\mu)]$, while the period of the oscillations is set by $\mu$.  
In the topological phase for S a MZM also emerges for  $\omega\approx0$ at the NS interface.\cite{PhysRevB.86.085408} Thus, the maximum value of the odd-frequency pairing at $x=0$  corresponds to the emergence of a MZM at the NS interface. However, surprisingly,  we do not find the $1/|\omega|$ dependence reported in previous works\cite{PhysRevLett.99.037005,Takagi18,tamura18} The $1/|\omega|$ odd-frequency dependence has been used as a signal of the creation of the MZM quasiparticle. Here we attribute the lack of this frequency divergence to the fact that the MZM is not fully localized at the NS interface, but shows a finite spread into N, see next section.
As frequency increases in the topological phase, the even-frequency component acquires a finite value with a linear frequency dependence below the gap $E_{1}$ and the odd-frequency term remains roughly constant albeit with a slight decay. Close to the gap $E_{1}$, however, the even-frequency term exhibits a faster increase and develops a resonance feature at the gap edge, while outside the gap  both components decay, see Fig.\,\ref{fig4}(c,e,g). 
In the trivial phase in S both even- and odd-frequency  magnitudes are finite for all $\omega$, although they acquire smaller values than in the topological phase, as seen in both Figs.\,\ref{fig3} and \,\ref{fig4}(d,f,h).
Both magnitudes develop a kink-like feature at the gap $\omega=\mu$. Outside this gap both components decay with $\omega$. Note that they also exhibit an exponential decay (approximately within $2\xi$) without oscillations since the wavevectors $q_{e,h}$ are purely imaginary for $\mu<0$.

 \begin{figure}[!t]
\centering
\includegraphics[width=.49\textwidth]{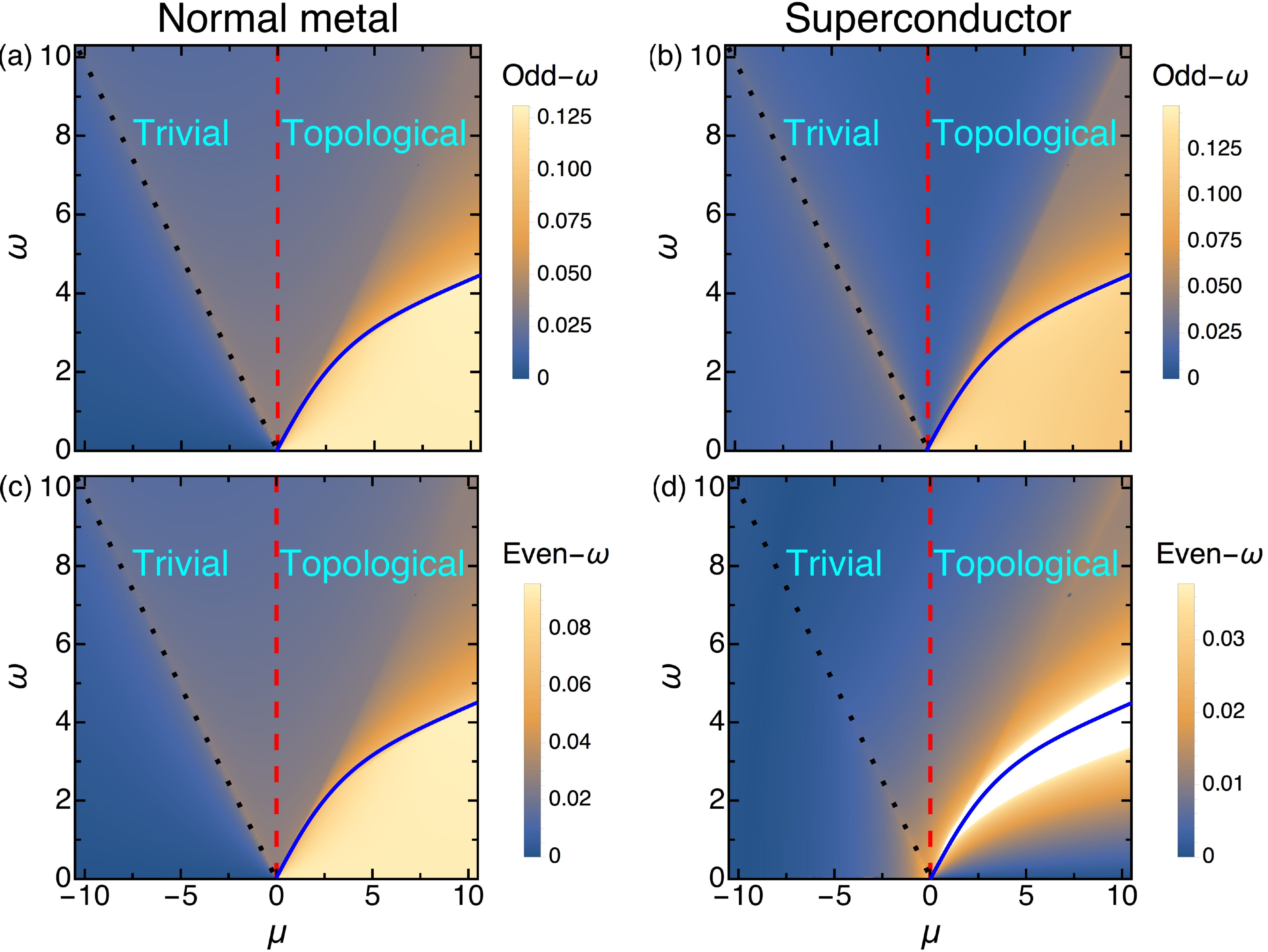} 
\caption{(Color online.) Odd- (a, b) and even (c, d) frequency pair magnitudes in the frequency ($\omega$) versus chemical potential ($\mu$) plane for a NS junction, where panels (a, c) correspond to N and  (b,d) to S. For S only the interface amplitude is plotted.
$\omega$ and $\mu$ are given in units of  $m\Delta^2=1$  for $m=0.5$, while $\Delta=\sqrt{2}$ and $V=0$. 
Here, $x'=0$ and  $x=\mp 0.14\xi$ away from the interface into the N and S regions, respectively. Red lines denote the topological phase transition at $\mu = 0$. The blue and black dotted lines denote the $\omega = E_1$ and $\omega=-\mu$ curves, respectively.}
\label{fig6}
\end{figure}

So far we have analyzed the pair amplitudes in N and S regions at fixed chemical potential, $\mu$, which is the parameter that tunes the trivial and topological phases in S. Now, in Fig.\,\ref{fig6} we present the pair magnitudes as a function of both frequency and chemical potential $\mu$. The topological and trivial phases  are separated by the red vertical line at  $\mu=0$. The results in both the N and S regions for both even- and odd-frequency are clearly separated according to the different expressions for the superconducting gap in different chemical potential regimes of the Kitaev model, indicated by the blue, black dashed and dotted lines. 
In N and S both pair magnitudes are larger in the topological phase than in the trivial phase. The odd-frequency component is clearly dominating in the whole subgap regime $\omega<E_1$ (solid blue curve) in the topological phase, although relatively large even-frequency exists for  frequencies around the gap  $\omega=E_1$. We also observe finite odd-frequency component in the trivial phase both in the S and N regions, with a slightly enhanced value around the gap, $\omega=-\mu$ (see dotted line).

\subsubsection{LDOS and local pairing}
After analyzing the pair amplitudes, we now discuss the LDOS in the superconducting region S and its relation with the  interface pair magnitudes.
We focus on the S region  because  the wavefunction of the predicted MZM emerges located at the NS interface and decays into the bulk of S,\cite{PhysRevB.86.085408} and therefore the LDOS in the S region will elucidate the relation, if any, between the odd-frequency pairing and the LDOS.

\begin{figure}[!t]
\centering
\includegraphics[width=.49\textwidth]{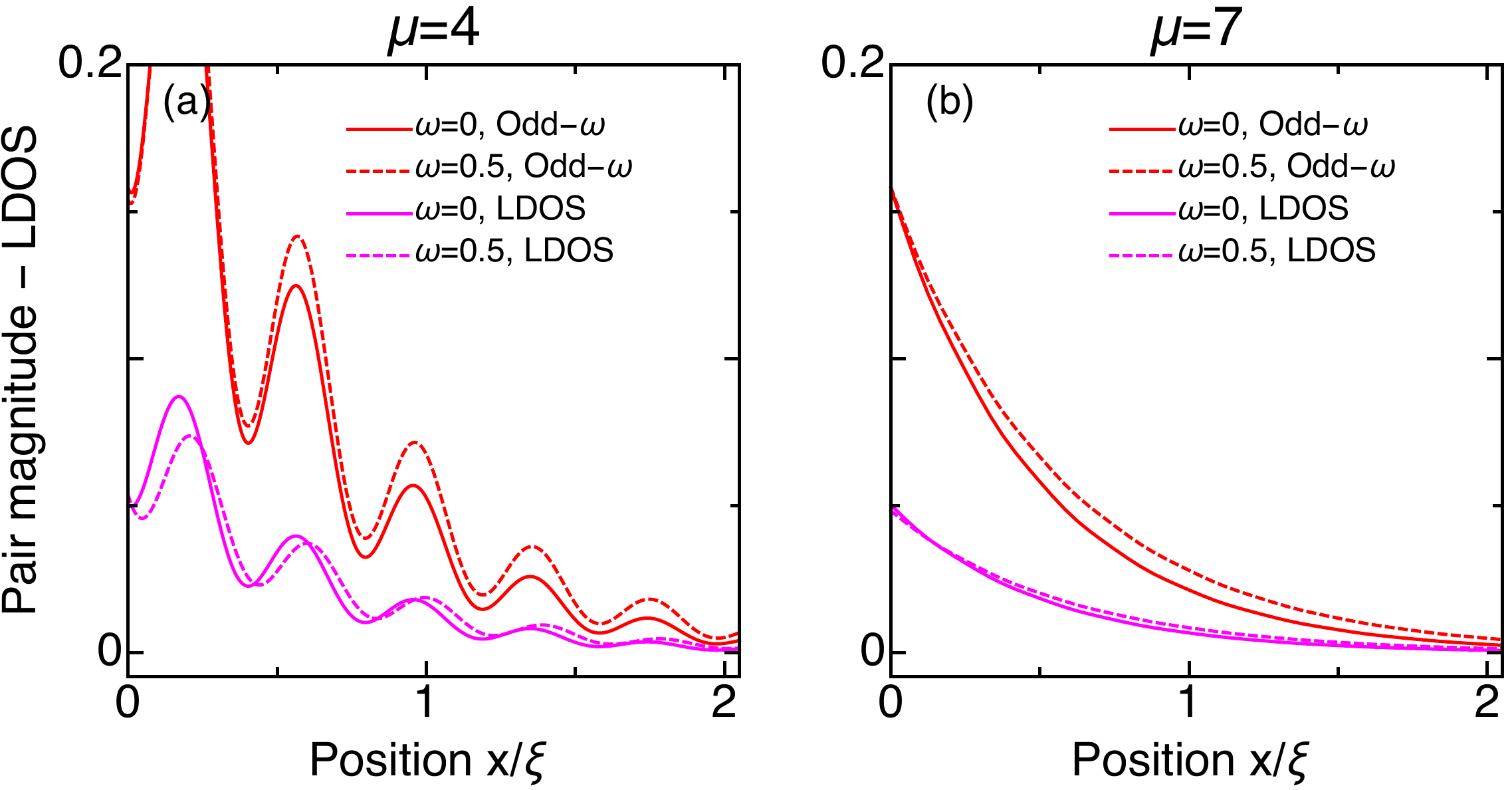} 
\caption{(Color online.) Spatial dependence of the interface LDOS (magenta) and magnitude of the local odd-frequency pair amplitude (red) for two chemical potentials $\mu$ in the topological phase. Solid and dashed curves correspond to $\omega\approx0$ and  $\omega\approx0.5$, respectively. Frequencies are given in units of $E_{1}$. Parameters same as in Fig.\,\ref{fig2}. 
}
\label{fig8}
\end{figure}

The electronic LDOS is obtained from $\rho(x,\omega)=-(1/\pi){\rm Im}G^{r}_{ee}(x,x,\omega)$, where $G_{ee}^{r}$ is the regular electron-electron Green's function whose  general form is given by,
\begin{equation}
\label{GeeLDOS}
\begin{split}
&G^{r,{\rm S}}_{ee} (x,x',\omega)= \\
&= -s  \Big[ U_{q_{h}}  e^{-iq_{h}|x-x'|} +  U_{q_{e}}  e^{iq_{e}|x-x'|} \Big] \\
&+ s  \Big[ U_{q_{h}}   r_{q_{h}q_{h}}  e^{-iq_{h}(x+x')} + U_{q_{e}}  r_{q_{e}q_{e}}  e^{iq_{e}(x+x')} \Big] \\
&- s  U_{q_{h}}   \frac{V_{q_{h}}}{V_{q_{e}}} r_{q_{e}q_{h}}  \Big[ e^{-iq_{h}x + i q_{e}x'} + e^{iq_{e}x - i q_{h}x'} \Big]\,,
\end{split}
\end{equation}
where the first, second, and third lines correspond to contributions from the bulk, normal, and Andreev reflections, respectively. The explicit derivation of $G_{ee}^{r,{\rm S}}$ in Eq.\,(\ref{GeeLDOS}) is given in Appendix \ref{AppNS}.
The first observation in Eq.\,(\ref{GeeLDOS}) is that it acquires a form very similar to the anomalous Green's function given by Eq.\,(\ref{GS_NS}), namely, it contains contributions from the bulk (first line) and interface (second and third lines), which suggests that the LDOS can be then written as $\rho=\rho_{\rm B}+\rho_{\rm I}$. We are thus interested in the interface LDOS as it exhibits the emergence of the MZM at the NS interface in the topological region. By using Eq.\,(\ref{GeeLDOS}) for subgap frequencies $\omega<E_{1}$ in the topological phase, we obtain
\begin{equation}
\label{LDOS_S}
\begin{split}
\rho_{\rm I}(x,\omega)&=-\frac{{\rm e}^{-2\kappa x}}{\pi}\times\\
{\rm Im}\Big\{
&
  s\Big[ U_{q_{h}}   r_{q_{h}q_{h}}  e^{-2ik_{\mu}x} 
  + U_{q_{e}}  r_{q_{e}q_{e}}  e^{2ik_{\mu}x}  \\
&- 2 U_{q_{h}}   \frac{V_{q_{h}}}{V_{q_{e}}} r_{q_{e}q_{h}}\Big]\Big\}\,,
\end{split}
\end{equation}
where we have used the fact that in the topological regime the wavevectors can be written as $q_{e,h}=k_{\rm F}\pm i\kappa$, with $\kappa=\sqrt{E_{1}^{2}-\omega^{2}}k_{\mu}/(2\mu)$ and $k_{\mu}=\sqrt{2m\mu_{S}/\hbar^{2}}$. 
Around $\omega\approx0$, Eq.\,(\ref{LDOS_S}) acquires the following simplified form
\begin{equation}
\label{LDOS_S2}
\begin{split}
\rho_{\rm I}(x,\omega\approx0)&\approx-\frac{2{\rm e}^{-2\kappa x}}{\pi}\times\\
&{\rm Im}\Big\{
is\Big[r_{q_{e}q_{h}}+r_{q_{h}q_{h}}{\rm sin}(2k_{\mu}x)\Big]
\Big\}\,.
\end{split}
\end{equation}
Thus, there is an exponential decay term, which multiplies the total interface LDOS. This is in agreement with the expectation for a MZM as such a state is present at the interface and exponentially decays towards the bulk of S, as illustrated in Fig.\,\ref{fig8}(a). Moreover, the exponential decay is accompanied by an oscillatory profile due to normal reflections, which, in general, coexists with Andreev reflections. Deep in the topological phase, however, normal reflections are very small and the oscillatory profile is not observed anymore, as shown in Fig.\,\ref{fig8}(b). 

Next, we analyze the possible correlation between the even- and odd-frequency pair amplitudes with the LDOS. First of all, the interface LDOS given by Eq.\,(\ref{LDOS_S2}) is set by both normal and Andreev reflections through the coefficients $r_{q_hq_h(q_eq_e)}$ and $r_{q_{e}q_{h}}$, respectively. By inspection of  the pair amplitudes in Eqs.\,(\ref{pairS}), we note that it is only the odd-frequency component that is dependent on both normal and Andreev scattering processes. Furthermore, since the LDOS is a local quantity, it is natural to analyze the local pair amplitude (i.e.~$x=x'$), where only the odd-frequency component survives and acquires the form
\begin{equation}
\begin{split}
f^{r, {\rm O}}_{\rm I}(x,\omega\approx0)&\approx-2s{\rm e}^{-2\kappa x}\times\\&\Big[r_{q_{e}q_{h}} +ir_{q_{h}q_{h}}{\rm sin}(2k_{\mu}x)\Big]\,.
\end{split}
\end{equation}
This expression exhibits very strong similarities with the interface LDOS in the topological phase given by Eq.\,(\ref{LDOS_S2}). 
Both the local odd-frequency pair amplitude and LDOS have an exponential decay and an oscillatory profile due to normal reflections with period set by $k_{\mu}$, as we also illustrate in Fig.\,\ref{fig8}. Interestingly, deep in the topological phase, normal reflections are very small\cite{PhysRevB.92.205424,0034-4885-63-10-202} and the only process that determines both the interface LDOS and odd-frequency  amplitude is the Andreev reflection. 
As a consequence, by measuring the Andreev reflection, both the interface LDOS and the odd-frequency pair amplitude can be simultaneously characterized in the topological phase.

\subsubsection{Local conductance and local pairing}
At this point it is  important to mention that  we can also correlate the local conductance across the NS junction with odd-frequency pairing. 
The zero-temperature single mode conductance for an incident quasielectron from S is given by  $G=(e^{2}/h)(1+|r_{q_{e}q_{h}}|^{2}-|r_{q_{e}q_{e}}|^{2})$. 
Since normal reflections can be negligibly small, for example in transparent junctions with equal chemical potential in N and S or by tuning the barrier transmission through $V$, the Andreev amplitude $r_{q_{e}q_{h}}$ can be accessed directly from the local conductance $G$. Moreover, in the topological phase at low frequencies the Andreev reflection is large due to the presence of MZMs, resulting in a large local conductance.\cite{PhysRevB.82.180516,PhysRevB.95.205439} 
At the same time the MZM induces large values of the odd-frequency pairing at low frequencies, discussed above and seen in Fig.\,\ref{fig4}(c). Hence, we conclude that the large values of local conductance $G$ at low frequency necessarily signals the presence of large odd-frequency pair amplitudes. The same analysis applies to N but then both even- and odd-frequency pair amplitudes are simultaneously determined.

We close this part by pointing out that experiments accessing the Andreev reflection have already been reported in LDOS and conductance measurements.\cite{chang15,doi:10.1063/1.4971394,gulonder,Gazibegovic17,zhang18} 
This thus shows that scattering processes at the NS interface represent a simple but powerful physical connection between odd-frequency pairing and  LDOS/conductance.


\subsection{Short SNS junctions}
So far we have shown the coexistence of even- and odd-frequency pairing in both the trivial and topological phases in NS junctions, as well as their relation to the scattering processes and also the LDOS. Next, we consider short SNS junctions, characterized by having  a normal region with vanishing length, $L_N \rightarrow 0$. We focus on the topological regime, namely $\mu >0$, as these junctions host a pair of MZM when the superconducting phase difference is $\phi = \pi$, and therefore this is an additional system in which to investigate the relationship between odd-frequency pairing and MZMs. The short junction limit guarantees that there are no higher energy levels within the junction which do not exhibit any Majorana character.

The pair amplitudes are obtained from the Green's functions, which are calculated  as for NS junctions by employing the BdG Hamiltonian in Eq.\,(\ref{eq1}).  The left region is replaced by a S region with $\Delta $, while the right region pairing potential is $\Delta e^{i\phi}$ and, in particular, we focus on the role of a finite phase difference.
The S regions are assumed to be semi-infinite and therefore there are no exterior edge effects, other than in the junction itself, located at $x=0$.  As in the NS case, we keep a finite barrier strength $V$ between the two parts of the junction. For more details on the derivation of the Green's functions, see Appendix \ref{AppSNS}.

The spectral properties of short SNS junctions based on Kitaev wires in the topological phase are already well understood.\cite{kitaev,PhysRevB.94.014502} As discussed before, a MZM emerges at each  end of the wire  in the topological phase, but in SNS junctions the superconducting phase difference $\phi$ also plays a crucial role. A topological short SNS junction hosts a pair of ABSs which disperse with $\phi$ and follow\cite{PhysRevB.53.9371,Kwon2004} $\omega_{\pm}=\pm E_{1}\sqrt{D}{\rm cos}(\phi/2)$, where $D$ is a parameter controlling the barrier transparency. These ABSs thus develop a crossing at $\omega=0$ for $\phi=\pi$ which is protected  by the conservation of the total fermion parity and signals the emergence of a pair of MZMs inside the junction.\cite{PhysRevB.53.9371,kitaev,Kwon2004} These ABSs naturally give very strong peaks in the subgap LDOS at the junction which depend on $\phi$.
Note that these topological ABSs are non-conventional as finite values of $D$ do not remove the crossing at $\phi=\pi$, a common effect found in trivial junctions.\cite{zagoskin,Beenakker:92,FURUSAKI1999809} 

The Green's functions in the left and right S regions capture the same information and therefore we only need to study the Green's function in the left S region, whose anomalous electron-hole component reads
\begin{equation}
\label{GFss}
\begin{split}
&G_{eh}^{r,{\rm S}}(x,x',\omega) =\\
&=s  \Big[ {\rm e}^{-i|x-x'|q_{h}} - {\rm e}^{i|x-x'|q_{e}}\Big]  {\rm sgn}(x-x') \\
&+ s \Big[r_{q_{e}q_{e}1}  {\rm e}^{-i(x+x')q_{e}} - r_{q_{h}q_{h}1}  {\rm e}^{i(x+x')q_{h}}\Big] \\
&+ s  \Big[ \frac{V_{q_{e}}U_{q_{e}}}{V_{q_{h}}U_{q_{h}}} r_{q_{h}q_{e}1} {\rm e}^{ i(q_{h} x'-q_{e}x)}
\\&-\frac{V_{q_{h}}U_{q_{h}}}{V_{q_{e}}U_{q_{e}}}  r_{q_{e}q_{h}1} {\rm e}^{  i(q_{h}x - q_{e} x')} \Big]\,,
\end{split}
\end{equation}
where 
the coefficients $r_{q_{e(h)}q_{e(h)}1}$ and $r_{q_{e(h)}q_{h(e)}1}$ correspond to the normal and Andreev reflection coefficients, respectively, which become phase dependent by picking up the phase $\phi$ from the right S and, therefore, are different from the coefficients for NS junctions. Here, the subscript 1 indicates that these coefficients correspond to processes in the left S, while $U_{i}$ and $V_{i}$ are the same as for NS junctions and phase-independent. Explicit expressions are given in Appendix \ref{AppSNS}. 

Despite the phase dependence of the scattering coefficients for SNS junctions, we find similarities to NS junctions in the overall structure of $G_{eh}^{r,{\rm S}}$.
In fact, in Eq.\,(\ref{GFss}) we  recognize the bulk (first square brackets) and interface (second and third square brackets) components, where the latter includes normal and Andreev reflections, respectively. 
Before going further, we point out that the emergence of ABSs correspond to poles, or zeroes in the denominator, of the Green's function, see e.g.~Ref.\,\onlinecite{zagoskin}. 
For the short SNS junction, the denominators of the normal and anomalous Green's functions coincide as they are both given by the denominator of the scattering coefficients $r_{q_{e(h)}q_{e(h)}1}$ and $r_{q_{e(h)}q_{h(e)}1}$.
Hence, in what follows we only need to  analyze  the interface component of the anomalous Green's function given by Eq.\,(\ref{GFss}), but can still pinpoint the energies of the ABSs. 

To proceed we employ Eqs.\,(\ref{EOF}) and find the even and odd-frequency interface pair amplitudes
\begin{equation}
\label{evenoddss} 
\begin{split}
f^{r,{\rm E}}_{\rm I} (x,x',\omega)&= I_{\rm AR}^{+}  \Big[ {\rm e}^{i(q_{h} x' -q_{e}x)} - {\rm e}^{ i(q_{h}x - q_{e} x')}\Big]\,,\\
f^{r,{\rm O}}_{\rm I} (x,x',\omega)&=I_{\rm AR}^{-}  
 \Big[ {\rm e}^{i(q_{h} x'-q_{e}x)} +  {\rm e}^{ i(q_{h}x - q_{e} x')}\Big]\\
&+  \Big[  I_{\rm NR}^{e} {\rm e}^{-i(x+x')q_{e}}- I_{\rm NR}^{h} {\rm e}^{i(x+x')q_{h}} \Big]\,,
\end{split}
\end{equation}
where $I_{\rm AR}^{\pm}(\phi,\omega)=\frac{s}{2}  \big[ \frac{U_{q_{e}} V_{q_{e}} }{U_{q_{h}} V_{q_{h}} }  r_{q_{h}q_{e}1} \pm \frac{U_{q_{h}} V_{q_{h}} }{U_{q_{e}} V_{q_{e}} }  r_{q_{e}q_{h}1} \big]$ and 
$I_{\rm NR}^{e(h)}=sr_{q_{e(h)}q_{e(h)}1}$ contain contributions from Andreev (AR) and normal (NR) reflections, respectively. Thus even- and odd-frequency pairings coexist and the effect enabling this coexistence is the Andreev reflection, through the coefficients $I_{\rm AR}^{\pm}$. 
Since normal reflection contributes only with even spatial parity, as seen in Eq.\,(\ref{GFss}), it only appears in the odd-frequency amplitude through coefficients, as for NS junctions.  The spatial behavior of Eqs.\,(\ref{evenoddss}) is in fact very similar to the S region of NS junctions.  

 \begin{figure}[!t]
\centering
\includegraphics[width=.49\textwidth]{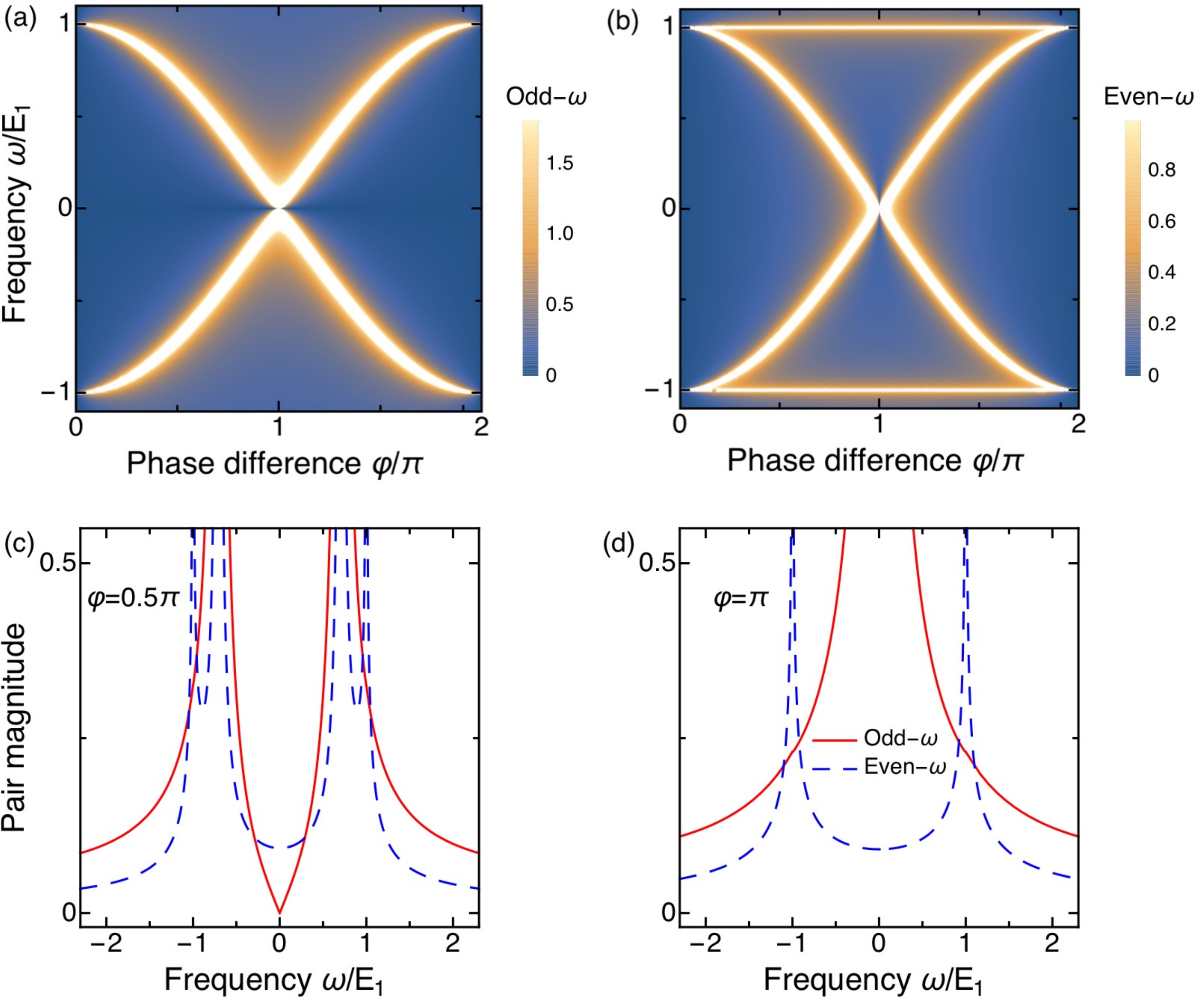} 
\caption{(Color online.) Magnitudes of the interface odd- (a) and even-frequency components (b) as a function of $\omega$ and $\phi$ for a short SNS junction and with their frequency dependence at $\phi=0.5\pi$ (c) and $\phi=\pi$ (d). At $\phi=\pi$ the odd-frequency amplitude exhibits a zero-frequency resonant peak, which stems from the protected zero-frequency MZM crossing.
We choose $x=-0.05\xi , x' = 0, V= 0, m=0.5, \Delta= 0.5$, and $E_{1} = 1$.}
\label{fig7}
\end{figure}

Further insight into the interface pairing is gained in Fig.\,\ref{fig7}(a,b), where we plot the even- and odd-frequency interface magnitudes from Eq.\,(\ref{evenoddss}) with respect to $\omega$ and $\phi$ in the left S region for finite $x-x'$, such that both amplitudes are finite. For a better visualization, in Fig.\,\ref{fig7}(c,d) we plot both amplitudes at $\phi=0.5\pi$  and $\phi=\pi$.
First we note that at $\phi=0$ both the odd- and even-frequency magnitudes capture the  gap edges through resonance peaks at $\omega=\pm E_{1}$, as seen in Fig.\,\ref{fig7}(a,b).
Around zero frequency at $\phi=0$  both  interface pair magnitudes vanish as $r_{q_{e(h)}q_{e(h)}1} = r_{q_{e(h)}q_{h(e)}1} =  0$ for a transparent junction $V=0$. We have verified (not shown) that   finite values of $V$ lead to finite scattering coefficients even at zero phase, giving rise to finite even- and odd-frequency amplitudes.

As discussed before, a finite phase difference $\phi$ drives the formation of subgap ABSs.
In fact, this is observed  in Fig.\,\ref{fig7} where both pair magnitudes  develop peaks with large magnitudes and cosine-like behavior in (a,b). These bright regions result from zeroes in the denominator of the scattering coefficients $r_{q_{e(h)}q_{e(h)}1}$ and $r_{q_{e(h)}q_{h(e)}1}$, which, as exposed above, reveal the formation of a pair of ABSs with energies lower than the gap $E_{1}$. This is in fact the main characteristic of the low-energy spectrum in short Kitaev-based SNS junctions.\cite{PhysRevB.53.9371,Kwon2004} 
We note that around the ABS energies, the odd-frequency magnitude is usually larger than the even-frequency magnitude, as seen in Fig.\,\ref{fig7}(c,d). 
At zero frequency at $\phi=0.5\pi$, however, even-frequency exhibits a finite value while  odd-frequency gets completely suppressed as it is an odd function. It is worth noting that at very low frequencies, the odd-frequency magnitude away from $\phi =\pi$ and the MZMs, roughly  develops a linear dependence on frequency $|f^{r,{\rm O}}_{\rm I}|\approx |\omega|$. This is seen in Fig.\,\ref{fig7}(c) and we have further verified that this statement holds for $0<\phi<\pi$.

Moreover, apart from ABSs, the even-frequency magnitude  also has a $\phi-$ independent branch at $\omega = \pm E_1$, corresponding to the gap edge. This can be understood from the  expression for the even-frequency amplitude, which gives rise to extra peaks through the factor $s$, since for $\omega =\pm E_1$ we have $U_{q_{e}} = U_{q_{h}}$ and thus the denominator of $s$ becomes zero. 
On the other hand, this peak does not appear in the odd-frequency as, for $\omega =\pm E_1$, we also have $V_{q_{e}} = V_{q_{h}}$, $r_{q_{e(h)}q_{e(h)}1}=0$ and $r_{q_{e}q_{h}1}=r_{q_{h}q_{e}1}$, thus rendering the whole expression in the numerator of $I_{\rm AR}^{-}$ zero.

Finally, focusing on the situation at $\phi = \pi$ in Fig.\,\ref{fig7}(d) we find that both pair magnitudes exhibit resonances, but with different features. For the odd-frequency amplitude the scattering processes interfere constructively and result in a resonance peak at $\omega=0$, which corresponds to the protected zero-frequency crossing in the ABS spectrum and therefore to the emergence of MZMs.\cite{PhysRevB.53.9371,Kwon2004} The resonance peak results in $|f^{r,{\rm O}}_{\rm I}|\approx 1/|\omega|$, in agreement with the appearance of the MZMs.
We however stress  that the odd-frequency amplitude is also finite for $\omega\neq0$ and $\phi\neq\pi$,  proving that the existence of odd-frequency pairing does not require the MZMs.
On the other hand, for even-frequency amplitude, the resonances interfere destructively and therefore no peak is present at low frequencies, although the $\phi-$ independent peaks remain at the gap edges $\omega =\pm E_1$.  Thus the MZM peak is only observed in the odd-frequency amplitude.
We note here that this destructive interference has its maximum for $\phi = \pi$. If we move slightly away from this value, the even-frequency magnitude exhibits larger values. 

In the end, we emphasize that the observation of the zero-frequency peak at $\phi=\pi$ is a clear signature of dominant odd-frequency pairing. As discussed at the beginning of this subsection, the zero-frequency peak in the LDOS at the junction arises because the topological ABSs develop a  crossing at $\phi=\pi$, protected  by the conservation of the total fermion parity, which then reflects as a zero-frequency peak in the LDOS at the junction.\cite{PhysRevB.53.9371,kitaev,Kwon2004}
Moreover, under these conditions, the topological ABSs are $4\pi$-periodic and lead to the $4\pi$-periodic fractional Josephson effect.\cite{PhysRevB.53.9371,kitaev,Kwon2004} Remarkably, the protected zero-frequency crossing at $\phi=\pi$ appears only in the odd-frequency pairing as a resonance peak, as  discussed in previous paragraph and shown in Fig.\,\ref{fig7}(d). Therefore, the observation of the zero-frequency peak in LDOS or fractional Josephson effect can both be seen as signatures of odd-frequency pairing. On the experimental side, recent advances in the measurement of ABSs\cite{PhysRevX.3.041034,Bocquillon17,Woerkom17,PhysRevLett.121.047001} and Josephson effect\cite{PhysRevLett.119.187704,Bocquillon2018,hart19,fourpi,ren19} indicates that the odd-frequency signatures suggested here are achievable.

\section{Conclusions}
\label{concl}
In this work we have implemented a scattering Green's function approach and analytically investigated the emergence of induced superconducting pair correlations in NS and short SNS Kitaev junctions. Our main motivation was to provide further conceptual understanding on the relationship between the topological phase in the Kitaev model, particularly its MZMs, and odd-frequency pairing. We also aimed at characterizing  pair amplitudes based on scattering processes, which hold  relevance in conductance experiments.
In general, we found that in both the trivial and topological phases there is a coexistence of  even- and odd-frequency pair amplitudes, which arise due to the spatial invariance breaking at interfaces. In particular, the existence of odd-frequency pairing does not require the presence of MZMs, although when MZMs emerge odd-frequency pairing is enhanced.

More specifically, we demonstrated that the Andreev reflection  acts as a spatial parity mixer and, therefore, allows for the coexistence of even- and odd-frequency pairing. Hence, despite Kitaev junctions being based on $p$-wave superconductors, the role of Andreev reflection in the generation of even- and odd-frequency correlations is still the same as in $s$-wave superconductor junctions.\cite{PhysRevB.92.205424,PhysRevB.96.155426,PhysRevB.98.075425} 
On the other hand, our results show that normal reflections  contribute only to the odd-frequency pairing in S, unlike in $s$-wave superconductor junctions.\cite{PhysRevB.92.205424,PhysRevB.98.075425} 
This is an  interesting conceptual result that adds further understanding to odd-frequency pairing in junctions and the characterization of pair amplitudes by means of normal and Andreev reflections. 
However, we found that in the S regions of NS junctions it is possible to characterize the pair amplitudes solely by means of Andreev reflections deep in the topological phase. 
Moreover, in this regime, the Andreev reflection can be obtained from LDOS or conductance measurements\cite{chang15,doi:10.1063/1.4971394,gulonder,Gazibegovic17,zhang18} and therefore such quantities can be used to determine the odd-frequency amplitudes. 
Importantly, we also showed that the odd-frequency component in the S region of NS junctions does not exhibit the $1/|\omega|$ dependence reported in previous works in the presence of an isolated MZM.\cite{lutchyn16,Takagi18,tamura18} This effect we attribute to the MZM having a finite spread beyond the junction.

In short SNS junctions we found that both even- and odd-frequency pair amplitudes capture the emergence of the topological Andreev bound states, the main characteristic in these Kitaev junctions and also clearly visible in the LDOS. While  for $0<\phi<\pi$  the odd-frequency pairing exhibits a linear frequency dependence ($\sim|\omega|$) at low frequencies, it develops a divergence ($\sim1/|\omega|$) at $\phi=\pi$, which signals the emergence of the MZMs. 
This is in stark contrast to NS junctions where, despite the presence of a MZM at the interface, the pair amplitude does not diverge at low-frequencies. 
On the possible experimental signatures in short SNS junctions, our calculations show that the zero-frequency peak in LDOS at $\phi=\pi$ in the junction and the $4\pi$-fractional Josephson effect can both be seen as signatures of dominant odd-frequency pairing.

\textit{Note Added}: While finishing the manuscript, two preprints, Refs.~\onlinecite{Takagi18,tamura18} appeared where odd-frequency in Kitaev wires was investigated, but with their main focus only on the topological phase in NS junctions. 

\section{Acknowledgements}
We thank L.~Komendova, C.~Triola, M. ~Mashkoori, P.~Dutta, F.~Parhizgar, and D.~Kuzmanovski for interesting discussions.
 This work was made possible by support from the Swedish Research Council (Vetenskapsr\aa det, 621-2014-3721), the G\"{o}ran Gustafsson Foundation, the Knut and Alice Wallenberg Foundation through the Wallenberg Academy Fellows program, and the European Research Council (ERC) under the European Union's Horizon 2020 research and innovation programme (ERC-2017-StG-757553).
 
\bibliography{biblio}

\appendix
\begin{widetext}
\section{Green's functions}
\label{App1}
In this Appendix we describe how we access the pair amplitudes. We construct the retarded Green's functions following Refs.\,\onlinecite{PhysRevB.96.155426,PhysRevB.98.075425} from scattering states with outgoing boundary conditions.\cite{PhysRev.175.559} They then read,
\begin{equation}
\label{GFUNCTION}
\begin{split}
&G^{r}(x,x',\omega)=
\begin{cases}
  a_{1} \Psi_{1}(x)\tilde{\Psi}_{3}^{T}(x')+a_{2} \Psi_{1}(x)\tilde{\Psi}_{4}^{T}(x') 
   + a_{3} \Psi_{2}(x)\tilde{\Psi}_{3}^{T}(x')+a_{4} \Psi_{2}(x)\tilde{\Psi}_{4}^{T}(x'), & x>x', \\
   b_{1} \Psi_{3}(x)\tilde{\Psi}_{1}^{T}(x')+ b_{2} \Psi_{4}(x)\tilde{\Psi}_{1}^{T}(x') 
   +   b_{3} \Psi_{3}(x)\tilde{\Psi}_{2}^{T}(x')+  b_{4} \Psi_{4}(x)\tilde{\Psi}_{2}^{T}(x'), & x<x',
\end{cases}
\end{split}
\end{equation}
where $\Psi_{i}$ correspond to the scattering processes at the interface and are obtained from the solutions to the Hamiltonian $H_{\rm BdG}$ in the specific region in question (see subsequent appendices), given by Eq.\,(\ref{eq1}) in the main text.
Here, $\tilde{\Psi}_{i}$ represent the conjugated processes obtained from $\tilde{H}_{\rm BdG}(p) = H^*_{\rm BdG} (-p) = H^T_{\rm BdG} (-p)$. Since spin is not explicitly present, $\Psi_{i}$ represent four scattering processes only: right moving particles (electrons and holes) from the left region and left moving particles from the right region. Notice that, since the scattering states are determined in each region, Eq.\,(\ref{GFUNCTION})  provides the Green's function in the left and right regions separately.

The coefficients $a_i,b_i$ in (\ref{GFUNCTION}) are calculated by integrating the equation of motion
\begin{equation}
\label{eomomega}
[\omega - H_{\rm BdG}(x)] G^r (x,x',\omega) = \delta (x-x')
\end{equation}
around $x=x'$ and are shown in subsequent appendices.   
The total Green's function is a $2\times2$ matrix in electron-hole space and reads
\begin{equation}
\label{GFform}
G^r(x,x',\omega) = 
\begin{pmatrix}
G^r_{ee}(x,x',\omega) & G^r_{eh}(x,x',\omega)  \\
G^r_{he}(x,x',\omega) & G^r_{hh}(x,x',\omega)  
\end{pmatrix},
\end{equation}
where all elements are scalars because spin is not an active degree of freedom. Thus, the off-diagonal elements fully determine the pair amplitudes and we can write Eq.\,(\ref{pairingF}) in the main text.

\section{NS junctions}
\label{AppNS}
In this Appendix we present details of the calculation of the Green's function for NS junctions with the interface at $x=0$. 
We  first solve for the eigenvalues and eigenvectors of $H_{\rm BdG}$ in the normal region $x<0$ with $\Delta=0$ and the superconducting region $x>0$ with $\Delta\,{\rm e}^{i\phi}\neq 0$, with $\phi$ being the superconducting phase.
In this case, there are four scattering processes at the interface, which read
\begin{equation}
\label{ScatPsi_NS}
\begin{split}
\Psi_{1}(x)&=
\begin{cases}
u_{e}e^{ik_ex} +
 r_{ee} u_{e} e^{-ik_ex} + r_{eh} u_{h}e^{ik_hx}, \quad  \quad &x<0 \\ 
 t_{eq_{e}}  U_{q_{e}}^+ e^{iq_{e} x} + t_{eq_{h}} U_{q_{h}}^- e^{-iq_{h} x}, \quad  \quad &x>0
\end{cases}\\
\Psi_{2}(x)&=
\begin{cases}
u_{h}e^{-ik_hx}+ r_{hh} u_{h}e^{ik_hx} +
 r_{he} u_{e} e^{-ik_ex} , \quad  \quad &x<0 \\ 
 t_{hq_{e}}  U_{q_{e}}^{+} e^{iq_{e} x} + t_{hq_{h}} U_{q_{h}}^{-} e^{-iq_{h} x}, \quad  \quad &x>0
\end{cases}\\
\Psi_{3}(x)&=
\begin{cases}
t_{q_{e}e}u_{e}e^{-ik_ex} +
 t_{q_{e}h} u_{h} e^{ik_hx}, \quad  \quad &x<0 \\ 
U_{q_{e}}^+ e^{iq_{e} x} +
 r_{q_{e}q_{e}}  U_{q_{e}}^+ e^{iq_{e} x} + r_{q_{e}q_{h}} U_{q_{h}}^- e^{-iq_{h} x}, \quad  \quad &x>0
\end{cases}\\
\Psi_{4}(x)&=
\begin{cases}
t_{q_{h}e}u_{e}e^{-ik_ex} +
 t_{q_{h}h} u_{h} e^{ik_hx}, \quad  \quad &x<0 \\ 
U_{q_{h}}^+ e^{iq_{h} x} +
 r_{q_{h}q_{e}}  U_{q_{e}}^+ e^{iq_{e} x} + r_{q_{h}q_{h}} U_{q_{h}}^- e^{-iq_{h} x}, \quad  \quad &x>0
\end{cases}
\end{split}
\end{equation}
where 
\begin{equation}
\label{eqcs}
\begin{split}
u_{e}&=A
\begin{pmatrix}
1\\
0
\end{pmatrix},\,
u_{h}=B
\begin{pmatrix}
0\\
1
\end{pmatrix},\\
U_{q_{e}}^-&=b\begin{pmatrix}
-e^{i\phi /2} U_{q_{e}} V_{q_{e}}\\
e^{-i\phi /2} V_{q_{e}} 
\end{pmatrix}, \,
U_{q_{h}}^-=c\begin{pmatrix}
-e^{i\phi /2} U_{q_{h}} V_{q_{h}}\\
e^{-i\phi /2} V_{q_{h}} 
\end{pmatrix}\,,
U_{q_{e}}^+=a\begin{pmatrix}
e^{i\phi /2} U_{q_{e}} V_{q_{e}}\\
e^{-i\phi /2} V_{q_{e}} 
\end{pmatrix},\,
U_{q_{h}}^+=d\begin{pmatrix}
e^{i\phi /2} U_{q_{h}} V_{q_{h}}\\
e^{-i\phi /2} V_{q_{h}} 
\end{pmatrix}\,,
\end{split}
\end{equation}
are the eigenvectors that determine the scattering processes given by Eqs.\,(\ref{ScatPsi_NS}), 
and
\begin{equation}
\begin{split}
U_{q_{e},q_{h}}  V_{q_{e},q_{h}}&=\frac{\omega+\epsilon_{q_{e},q_{h}}}{ \Delta q_{e,h}}  \sqrt{\frac{\omega-\epsilon_{q_{e},q_{h}}}{2\omega}}\,,\quad
V_{q_{e},q_{h}}=\sqrt{\frac{\omega-\epsilon_{q_{e},q_{h}}}{2\omega}} \,,\quad
\epsilon_{q_{e,h}}= \frac{h^{2}q_{e,h}^{2}}{2m}-\mu\,.
\end{split}
\end{equation}
define the coherence factors $U_{i}$ and $V_{i}$. The coefficients $A,B,a,b,c,d$ in Eqs.\,(\ref{eqcs}) are included in order to normalize the respective eigenvectors. Although these coefficients  are important when calculating e.g.~conductance, we do not need to define them because it turns out that they do not appear in the final expressions for the Green's functions. 
The  scattering process $\Psi_{1}$ in Eqs.\,(\ref{ScatPsi_NS}) represents the following process: an incoming (right moving) electron from N with wavefunction $u_{e}e^{ik_{e}x}$ can be reflected into N either as a left moving electron, with wavefunction $u_{e}e^{-ik_{e}x}$ and amplitude $r_{ee}$, or a left moving hole, with wavefunction $u_{h}e^{ik_h{x}} $and amplitude $r_{eh}$, or transmitted into  S  as a right moving quasielectron, with wavefunction $U^{+}_{q_{e}}e^{iq_{e}x}$ and amplitude $t_{eq_{e}}$, or as a right moving quasihole, with wavefunction $U^{-}_{q_{h}}e^{-iq_{h}x}$ and amplitude $t_{eq_{h}}$. Likewise, $\Psi_{2}$ describes a right moving hole from N towards the NS interface. 
On the other hand,   $\Psi_{3(4)}$ describe left moving electron (hole) from S towards the NS interface. The reflection process of an electron (hole) into a hole (electron) is known as Andreev reflection. The same applies for the quasielectrons and quasiholes in the S region. The wavevectors $k_{e(h)}$ that appear in Eqs.\,(\ref{ScatPsi_NS}) correspond to electrons (holes) in N and $q_{e(h)}$ to quasielectrons (quasiholes) in S and are defined by Eqs.\,(\ref{wavevector}) in the main text.
They can become complex  depending of the values of $\omega$, $\mu$ and $\Delta$ and is detailed account for the values for the latter are presented in Table \ref{momentaexpressions}.

 \begin{table}[!t]
 	\begin{center}
	\resizebox{0.43\columnwidth}{!}{%
        \begin{tabular}{|c|c|c|}
        	\hline
        	$\mu$ & $\omega$ & $q_{e}$, $q_{h}$ \\
        	\hline
        	\multirow{3}{*}{$\mu \geq m \Delta^2$} & $\omega\geq |\mu|$ & \makecell{$q_{e} = \sqrt{2m} \Big( \mu - m\Delta^2 + \sqrt{\omega^2 - E_1^2}\Big)^{1/2}$ \\   $q_{h} = i\sqrt{2m} \Big(-\mu + m\Delta^2 + \sqrt{\omega^2 - E_1^2}\Big)^{1/2}$} \\
        	\cline{2-3}
        	&$E_1 \leq \omega \leq |\mu|$ & $q_{e,h} = \sqrt{2m} \Big(\mu - m\Delta^2 \pm \sqrt{\omega^2 - E_1^2}\Big)^{1/2}$    \\
        	\cline{2-3}
        	& $0 \leq \omega \leq E_1 $  & $q_{e,h} = \sqrt{2m} \Big(\mu - m\Delta^2 \pm i\sqrt{E_1^2 - \omega^2}\Big)^{1/2}$   \\
        	\hline
        	\multirow{3}{*}{$\frac{m \Delta2}{2} \leq \mu \leq m \Delta^2$} & $\omega\geq |\mu|$ & \makecell{$q_{e} = \sqrt{2m} \Big( \mu - m\Delta^2 + \sqrt{\omega^2 - E_1^2}\Big)^{1/2}$ \\   $q_{h} = i\sqrt{2m} \Big(-\mu + m\Delta^2 + \sqrt{\omega^2 - E_1^2}\Big)^{1/2}$}   \\
        	\cline{2-3}
        	&$E_1 \leq \omega \leq |\mu|$ & $q_{e,h} = i\sqrt{2m} \Big(-\mu + m\Delta^2 \mp \sqrt{\omega^2 - E_1^2}\Big)^{1/2}$  \\
        	\cline{2-3}
        	&$0 \leq \omega \leq E_1 $  & $q_{e,h} = \sqrt{2m} \Big(\mu - m\Delta^2 \pm i\sqrt{E_1^2 - \omega^2}\Big)^{1/2}$  \\
        	\hline
        	\multirow{2}{*}{$\mu \leq \frac{m\Delta^2}{2}$} & $\omega\geq |\mu|$ & \makecell{$q_{e} = \sqrt{2m} \Big( \mu - m\Delta^2 + \sqrt{\omega^2 - E_1^2}\Big)^{1/2}$ \\   $q_{h} = i\sqrt{2m} \Big(-\mu + m\Delta^2 + \sqrt{\omega^2 - E_1^2}\Big)^{1/2}$} \\
        	\cline{2-3}
        	&$0 \leq \omega \leq |\mu|$ & $q_{e,h} = i\sqrt{2m} \Big(-\mu + m\Delta^2 \mp \sqrt{\omega^2 - E_1^2}\Big)^{1/2}$   \\
        	\hline
        \end{tabular}}
	\end{center}
\caption{Expressions for the momenta $q_{e}$, $q_{h}$ in S for different values of $\omega$, $\mu$, and $E_{1}=\sqrt{[2m\Delta^{2}/\hbar^{2}][\mu-m\Delta^{2}/(2\hbar^{2})]}$.}
\label{momentaexpressions}
 \end{table}

In order to fully determine the scattering states given by Eqs.\,(\ref{ScatPsi_NS}), the coefficients $r_{xx}$ and $t_{xx}$ need to be found. We thus integrate the BdG equations given by $H_{\rm BdG}\Psi(x)=E\Psi(x)$ for  NS junctions around $x=0$ and obtain
\begin{equation}
\label{NSSchrmatching}
\begin{split}
\Psi(0^+) &= \Psi (0^-)\,,\quad
[\partial_{x}\Psi(x>0) - \partial_{x}\Psi (x<0) ]_{x=0} = [\eta_1 \tau_{0} +  \eta_2 \tau_{y}]  \Psi(0), 
\end{split}
\end{equation}
where 
$\eta_{1}=2mV_{0}/\hbar^{2}$ and $\eta_{2}=m\Delta /\hbar$. Notice that the momentum dependence of the pairing potential in S, due to its $p$-wave nature, introduces a finite barrier strength through $\eta_{2}$. 
With this the coefficients read,
\begin{equation}
\label{eq:30}
\begin{split}
r_{ee} &=\frac{ ((\eta_1 + i (k_e - q_{e})) (\eta_1 + i (k_h + q_{h})) U_{q_{e}} + (\eta_1 + 
	i (k_h - q_{e})) (\eta_1 + i (k_e + q_{h})) U_{q_{h}})}
{ \Lambda}\\ 
&+ 
        \frac{(-\Delta^2 \eta_2^2 (U_{q_{e}} + U_{q_{h}}) - \Delta \eta_2 (q_{e} + q_{h}) (-1 + U_{q_{e}} U_{q_{h}}))}
       {\Lambda}\,, \\
 r_{hh} &= \frac{ (-(i \eta_1 + k_e + q_{e}) (i \eta_1 + k_h - q_{h}) U_{q_{e}} - (i \eta_1 + k_h + q_{e}) (i \eta_1 + k_e - q_{h}) U_{q_{h}})}{ \Lambda} \\ 
&+ \frac{(-\Delta^2 \eta_2^2 (U_{q_{e}} + U_{q_{h}}) - \Delta \eta_2 (q_{e} + q_{h}) (-1 + U_{q_{e}} U_{q_{h}}))}{ \Lambda}\,, \\      
       r_{q_{e}q_{e}} &= \frac{(-(\eta_1 - i (k_e - q_{e})) (\eta_1 + i (k_h + q_{h})) U_{q_{e}} +(\eta_1 + i (k_h + q_{e})) (\eta_1 - i (k_e - q_{h})) U_{q_{h}})}{ \Lambda}  \\
&+ \frac{(\Delta^2 \eta_2^2 (U_{q_{e}} - U_{q_{h}}) -\Delta \eta_2 (q_{e} - q_{h}) (1 + U_{q_{e}} U_{q_{h}}))}{ \Lambda}\,, \\
       r_{q_{h}q_{h}} &= \frac{((\eta_1 - i (k_e + q_{e})) (\eta_1 + i (k_h - q_{h})) U_{q_{e}} - (\eta_1 + 
	i (k_h - q_{e})) (\eta_1 - i (k_e + q_{h})) U_{q_{h}})}{ \Lambda }  \\ 
&+\frac{\Delta^2 \eta_2^2 (-U_{q_{e}} + U_{q_{h}}) + \Delta \eta_2 (q_{e} - q_{h}) (1 + U_{q_{e}} U_{q_{h}})}{\Lambda}\,,\\    
r_{eh} &= - \frac{2   k_e (q_{e} + q_{h} - \Delta \eta_2 (U_{q_{e}} + U_{q_{h}}))}
{\Lambda},\quad r_{he} = \frac{2  k_h [(q_{e} +q_{h}) U_{q_{e}} U_{q_{h}} + \Delta \eta_2 (U_{q_{e}} + U_{q_{h}})]}{\Lambda}\,, \\
r_{q_{e}q_{h}} &= \frac{2  [-\Delta^2 \eta_2^2 U_{q_{e}} + ((\eta_1 - i k_e) (\eta_1 + i k_h) + q_{e}^2) U_{q_{e}} -
	\Delta \eta_2 q_{e} (-1 + U_{q_{e}}^2)] V_{q_{e}}}{ \Lambda V_{q_{h}}}\,, \\
r_{q_{h}q_{e}} &= \frac{2  [-\Delta^2 \eta_2^2 U_{q_{h}} + ((\eta_1 - i k_e) (\eta_1 + i k_h) + q_{h}^2) U_{q_{h}} -
	\Delta \eta_2 q_{h} (-1 + U_{q_{h}}^2)] V_{q_{h}}}{ \Lambda V_{q_{e}}}\,, \\
t_{eq_{e}} &= \frac{2 i  k_e (\eta_1 + i (k_h + q_{h} - \Delta \eta_2 U_{q_{h}}))}
{\Lambda V_{q_{e}}},\quad
t_{eqh} = \frac{2   k_e (-i \eta_1 + k_h - q_{e} + \Delta \eta_2 U_{q_{e}})}
{ \Lambda V_{q_{h}}}\,,\\
t_{hq_{e}} &= - \frac{2 i   k_h (i \Delta \eta_2 + (\eta_1 - i (k_e - q_{h})) U_{q_{h}})}
{ \Lambda V_{q_{e}}},\quad
t_{hqh} = - \frac{2   k_h (\Delta \eta_2 + (i \eta_1 + k_e + q_{e}) U_{q_{e}})}{ \Lambda V_{q_{h}}}\,,\\
t_{q_{e}e} &= \frac{2   \{U_{q_{e}} [\Delta \eta_2 q_{h} + 
	q_{e} (-i \eta_1 + k_h + 
	q_{h}) U_{q_{e}}] -
[\Delta \eta_2 q_{e} + (q_{e}^2 + (-i \eta_1 + 
	k_h) q_{h}) U_{q_{e}}] U_{q_{h}}\} V_{q_{e}}}{ \Lambda}\,, \\
t_{q_{e}h} &= \frac{2   \{q_{e}^2 U_{q_{e}} - \Delta \eta_2 q_{e} U_{q_{e}}^2 + 
	q_{e} (i \eta_1 + k_e - q_{h}) U_{q_{h}}+ 
q_{h} U_{q_{e}} (-i \eta_1 - k_e + \Delta \eta_2 U_{q_{h}})\} V_{q_{e}}}{ \Lambda}\,,\\
t_{q_{h}e} &= \frac{2   \{\Delta \eta_2 (q_{h} U_{q_{e}} - q_{e} U_{q_{h}}) + 
	U_{q_{h}} [(I \eta q_{e} - k_h q_{e} + q_{h}^2) U_{q_{e}} + (-i \eta_1 + k_h - 
	q_{e}) q_{h} U_{q_{h}}]\} V_{q_{h}}}{ \Lambda}\,, \\
t_{q_{h}h} &= - \frac{2   \{(i \eta_1 + k_e + q_{e}) q_{h} U_{q_{e}} - [q_{h}^2 + 
	q_{e} (i \eta_1 + k_e + \Delta \eta_2 U_{q_{e}})] U_{q_{h}} + 
	\Delta \eta_2 q_{h} U_{q_{h}}^2\} V_{q_{h}}}{\Lambda}\,,
\end{split}
\end{equation}
where 
\begin{align}
\label{eq:25}
\begin{split}
\Lambda =& -(i \eta_1 + k_e + q_{e}) (-i \eta_1 + k_h + q_{h}) U_{q_{e}}
 + (i \eta_1 - k_h + 
q_{e}) (i \eta_1 + k_e - q_{h}) U_{q_{h}} \\
&+ \Delta^2 \eta_2^2 (U_{q_{e}} + U_{q_{h}}) + 
\Delta \eta_2 (q_{e} + q_{h}) (-1 + U_{q_{e}} U_{q_{h}}).
\end{split}
\end{align}

Furthermore, for the evaluation of the Green's function in Eq.\,(\ref{GFUNCTION}) we need the conjugated processes, $\tilde{\Psi}_{i}$, obtained from the conjugated Hamiltonian $\tilde{H}_{\rm BdG}(p) = H^*_{\rm BdG} (-p) = H^T_{\rm BdG} (-p) $. In matrix notation, the conjugated Hamiltonian, assuming a finite superconducting phase in $\Delta$ reads,
\begin{equation}
\label{eq:48}
\tilde{H}_{BdG}(p) = \left(
\begin{array}{cc}
\epsilon & -\Delta^* p  \\
-\Delta p & - \epsilon  
\end{array} \right) =
\left(
\begin{array}{cc}
\epsilon & -\Delta e^{-i\phi} p  \\
-\Delta e^{i\phi} p & - \epsilon  
\end{array} \right) =\left(
\begin{array}{cc}
\epsilon & \Delta e^{i\theta} p  \\
\Delta e^{-i\theta} p & - \epsilon  
\end{array} \right),
\end{equation}
where we have absorbed the minus sign in the new phase $\theta= -(\pi + \phi)$. Here the last expression has the same structure as $H_{\rm BdG}$ in Eq.\,(\ref{eq1}).  Thus, all our results from the normal processes $\Psi_{i}$ are valid for the conjugated ones, as long as we replace $\phi \rightarrow \theta$. Of course, for NS junctions, the phase $\phi$ does not play any role, and therefore we only need to track of the minus sign introduced by $\theta= -\pi$ in this case.
Using $\Psi_{i}$ and $\tilde{\Psi}_i$ derived above we next provide details on the Green's functions in N and S. 

\subsection{Green's function in N}
The Green's function in the normal region N is found by plugging the scattering states from Eqs.\,(\ref{ScatPsi_NS}) for $x<0$ into Eqs.\,(\ref{GFUNCTION}), with  $r_{xx}$ and $t_{xx}$ found from Eqs.\,(\ref{eq:30}). After some manipulations we obtain
\begin{equation}
\begin{split}
G_{ee}^{r}(x,x',\omega)&=  -\frac{i \eta}{2 k_e} \Big[  e^{i k_e  |x-x'|} + r_{ee} e^{-i k_e (x+x')}  \Big]  \,,\quad 
G_{hh}^{r}(x,x',\omega)= -\frac{i \eta}{2 k_h} \Big[e^{-i k_h  |x-x'|} + r_{hh} e^{i k_h (x+x')}  \Big] \,,
\\
G_{eh}^{r}(x,x',\omega)&=\frac{i \eta }{2k_e} r_{eh}  e^{-i(k_e x - k_h x')} \,,\quad
G_{he}^{r}(x,x',\omega)=- \frac{i \eta }{2k_e}  r_{eh}  e^{i(k_h x - k_e x')} \,,\\
\end{split}
\end{equation}
where $r_{eh}$ and $r_{ee,hh}$ correspond to the Andreev and normal reflection coefficients given by Eqs.\,(\ref{eq:30}). $G_{eh}^{r}$ can now be used to derive the even- and odd-frequency components for the N region, given by Eqs.\,(\ref{GehN_NS})  in the main text.

\subsection{Green's function in S}
In the superconducting region we proceed similarly as for the N region and obtain the following expressions for the elements of the Green's function,
\begin{equation}
\label{NS_APP_S}
\begin{split}
G_{ee}^{r}(x,x',\omega)&=  -s  \Big[ U_{q_{h}}  e^{-iq_{h}|x-x'|} +  U_{q_{e}}  e^{iq_{e}|x-x'|} \Big] + s  \Big[ U_{q_{h}}   r_{q_{h}q_{h}}  e^{-iq_{h}(x+x')} + U_{q_{e}}  r_{q_{e}q_{e}}  e^{iq_{e}(x+x')} \Big] \\
&- s  U_{q_{h}}   \frac{V_{q_{h}}}{V_{q_{e}}} r_{q_{e}q_{h}}  \Big[ e^{i(-q_{h}x +  q_{e}x')} + e^{i(q_{e}x - q_{h}x')} \Big]\,,\\
G_{eh}^{r}(x,x',\omega)&= s  \Big[ {\rm e}^{-i|x-x'|q_{h}} - {\rm e}^{i|x-x'|q_{e}} \Big]  {\rm sgn}(x-x') 
+ s \Big[ r_{q_{h}q_{h}}  {\rm e}^{-i(x+x')q_{h}} - r_{q_{e}q_{e}}  {\rm e}^{i(x+x')q_{e}} \Big] \\
&+ s  \frac{V_{q_{h}}}{V_{q_{e}}U_{q_{e}}}  r_{q_{e}q_{h}} \Big[ U_{q_{h}} {\rm e}^{ i(-q_{h}x + q_{e} x')} -  U_{q_{e}} {\rm e}^{ i(q_{e}x - q_{h} x')}\Big]\,,\\
G_{he}^{r}(x,x',\omega)&= 
s  \Big[ {\rm e}^{-i|x-x'|q_{h}} - {\rm e}^{i|x-x'|q_{e}} \Big]  {\rm sgn}(x-x') 
- s \Big[ r_{q_{h}q_{h}}  {\rm e}^{-i(x+x')q_{h}} - r_{q_{e}q_{e}}  {\rm e}^{i(x+x')q_{e}} \Big] \\
&- s  \frac{V_{q_{h}}}{V_{q_{e}}U_{q_{e}}}  r_{q_{e}q_{h}} \Big[ U_{q_{h}} {\rm e}^{ i(-q_{h}x' + q_{e} x)} -  U_{q_{e}} {\rm e}^{ i(q_{e}x' - q_{h} x)}\Big]\,,\\
G_{hh}^{r}(x,x',\omega)&= 
-s  \Big[ \frac{1}{U_{q_{h}}}  e^{-iq_{h}|x-x'|} +  \frac{1}{U_{q_{e}}}  e^{iq_{e}|x-x'|} \Big] - s  \Big[ \frac{1}{U_{q_{h}}}   r_{q_{h}q_{h}}  e^{-iq_{h}(x+x')} + \frac{1}{U_{q_{e}}}  r_{q_{e}q_{e}}  e^{iq_{e}(x+x')} \Big] \\
&- s  \frac{V_{q_{h}}}{V_{q_{e}}U_{q_{e}}} r_{q_{e}q_{h}}  \Big[ e^{i(q_{e}x-q_{h}x')} 
+ e^{i(q_{e}x' - q_{h}x)} \Big]\,,
\end{split}
\end{equation}
where $s=[i\eta (1+U_{q_{e}}^2)U_{q_{h}}]/[2 q_{h} (U_{q_{e}}^2 - U_{q_{h}}^2)]$ and  $r_{q_{e}q_{h}}$ and $r_{q_{e}q_{e},q_{h}q_{h}}$ correspond to the Andreev and normal reflection coefficients given by Eqs.\,(\ref{eq:30}). Here the anomalous electron-hole component  $G^{r}_{eh}$ allows for the calculation of  the induced pair correlations in the S region presented in the main text in Eqs.\,(\ref{GS_NS}) and (\ref{soddeven}). 
On the other hand, by using the regular component $G^{r}_{ee}$, we are able to write down Eq.\,(\ref{GeeLDOS}) in the main text and then calculate the LDOS.
The expressions given by Eqs.\,(\ref{NS_APP_S}) have terms which are not space dependent at $x=x'$ and terms that contain scattering coefficients. The former corresponds to bulk contributions, while the latter are terms generated by the interface. It is interesting to note that the bulk contribution in the anomalous term is odd under the exchange of spatial coordinates due to the $p$-wave nature of the Kitaev model.

\section{Short SNS junctions}
\label{AppSNS}
In this section we give further details on the calculations for short SNS junctions. In short junctions the length of the normal region is assumed to be extremely short, namely, $L_{\rm N}\rightarrow0$. Hence, a short SNS junction behaves as a Superconductor-Superconductor (SS') junction with a single interface and here we assume it to be located at $x=0$. Without loss of generality we consider the left superconductor described by $H_{\rm BdG}$ from Eq.\,(\ref{eq1}) with $\Delta (x)=\Delta$ for ($x<0$), while we use $\Delta (x)=\Delta\,{\rm e}^{i\phi}$ for the right superconductor ($x>0$), where $\phi$ is the superconducting phase difference.

We again proceed as explained above, following the same steps as for NS junctions.
First we calculate the quasiparticles eigenstates, then we construct the scattering wavefunctions and calculate the scattering coefficients.
As before, there are four scattering processes at the SS' interface and they read,
\begin{equation}
\label{scatqelss}
\begin{split}
\Psi_{1}(x) &= 
\begin{cases}
u_{q_{e1}}^+ e^{iq_{e}x} + r_{q_{e}q_{e}1}  u_{q_{e1}}^- e^{-iq_{e}x} + r_{q_{e}q_{h}1}  u_{q_{h1}}^+ e^{iq_{h}x}, \quad &x<0 \\
t_{q_{e}q_{e}2}  u_{q_{e2}}^+ e^{iq_{e}x} + t_{q_{e}q_{h}2}  u_{q_{h2}}^- e^{-iq_{h}x}, \qquad &x>0
\end{cases}\,,\\
\Psi_{2}(x) &= 
\begin{cases}
 u_{q_{h1}}^- e^{-iq_{h}x} + r_{q_{h}q_{e}1}  u_{q_{e1}}^- e^{-iq_{e}x} + r_{q_{h}q_{h}1}  u_{q_{h1}}^+ e^{iq_{h}x}, \quad &x<0 \\
t_{q_{h}q_{e}2}  u_{q_{e2}}^+ e^{iq_{e}x} + t_{q_{h}q_{h}2}  u_{q_{h2}}^- e^{-iq_{h}x}, \qquad &x>0
\end{cases} \\
\Psi_{3}(x) &= 
\begin{cases}
t_{q_{e}q_{e}1}  u_{q_{e1}}^- e^{-iq_{e}x} + t_{q_{e}q_{h}1}  u_{q_{h1}}^+ e^{iq_{h}x}, \quad &x<0 \\
 u_{q_{e2}}^- e^{-iq_{e}x} + r_{q_{e}q_{e}2}  u_{q_{e2}}^+ e^{iq_{e}x} + r_{q_{e}q_{h}2}  u_{q_{h2}}^- e^{-iq_{h}x}, \qquad &x>0
\end{cases} \\
\Psi_{4}(x) &= 
\begin{cases}
t_{q_{h}q_{e}1}  u_{q_{e1}}^- e^{-iq_{e}x} + t_{q_{h}q_{h}1}  u_{q_{h1}}^+ e^{iq_{h}x}, \quad &x<0 \\
 u_{q_{h2}}^+ e^{iq_{h}x} + r_{q_{h}q_{e}2}  u_{q_{e2}}^+ e^{iq_{e}x} + r_{q_{h}q_{h}2}  u_{q_{h2}}^- e^{-iq_{h}x}, \qquad &x>0
\end{cases}
\end{split}
\end{equation}
where the subscripts $1,2$ in the superconducting eigenvectors $u$ and the scattering coefficients $r_{xx}$, $t_{xx}$ differentiate between left or right S regions. The eigenvectors $u$ are the same as for the S region in NS junctions given by Eqs.\,(\ref{eqcs}) with $\phi=0$ ($\phi\neq0$) for the left (right) S. The meaning of $\Psi_{1,2,3,4}$ also follows from the NS junctions and corresponds to a right moving quasielectron  from left S, right moving quasihole from left S, left moving quasielectron from right S, left moving quasihole from right S, respectively.
Next, by integrating the BdG equations around the single interface $x=0$ for the short SNS junction, we arrive at two conditions that allow us to find the coefficients in the scattering wavefunctions $\Psi_{i}$,
\begin{align}
\label{matchingss}
\begin{split}
\Psi(0^+) =& \Psi(0^-), \\
\Psi'(0^+) - \Psi' (0^-)  = \eta_1 \Psi(0) +& i \Delta \eta_2 \left(
\begin{array}{cc}
0 & 1- e^{i\phi}  \\
-1+e^{-i\phi} & 0 
\end{array} \right) \Psi(0)\,.
\end{split}
\end{align}
where $\eta_{1}=2mV_{0}/\hbar^{2}$ and $\eta_{2}=m\Delta /\hbar$. Note that these boundary conditions arise from the single interface of the SS' junction located at $x=0$. Finite values of the normal region length introduce an additional set of equations to be evaluated at the position of the second interface. By solving these two equations we obtain all the coefficients. The expressions are long and we list them next. 

\subsection{Scattering coefficients}
After solving the system of equations given by Eqs.\,(\ref{matchingss}), with the scattering states $\Psi_{i}$ given by  Eqs.\,(\ref{scatqelss}), we obtain the following expressions for the scattering amplitudes for a right moving quasielectron from left S
\begin{align}
\label{eqsns4}
\begin{split}
r_ {q_{e}q_{e}1} &=  \{(q_ {e}^2 - q_ {h}^2) U_ {q_{e}} U_ {q_{e}} - 
e^{2 i\phi} (q_ {e}^2 - q_ {h}^2) U_ {q_{e}} U_ {q_{e}} + 
\Delta_ 0^2 (-1 + e^{i\phi})^2 \eta_ 2^2 (U_ {q_{e}}^2 - 
U_ {q_{e}}^2)  \\
&+e^{i\phi} \eta_ 1 (\eta_ 1 + 2 i q_ {h}) (U_ {q_{e}}^2 - 
U_ {q_{e}}^2) + 
\Delta_ 0 (-1 + 
e^{i\phi}) \eta_ 2 
\big[q_ {h} ((-1 + 
e^{i\phi}) U_ {q_{e}} + 
 (-1 + e^{i\phi}) U_ {q_{e}}^2 U_ {q_{e}} \\
& - (1 + e^{i\phi}) U_ {q_{e}} (-1 + 
U_ {q_{e}}^2)) +  q_ {e} (-(1 + 
e^{i\phi}) U_ {q_{e}} + (1 + 
e^{i\phi}) U_ {q_{e}}^2 U_ {q_{e}} - (-1 + 
e^{i\phi}) U_ {q_{e}} (1 + 
U_ {q_{e}}^2))\big]\}/\Lambda_ {ss}, \\
r_{q_{e}q_{h}1} &= -\big\{2 ( \Delta^2 (-1 + 
e^{i\phi})^2 \eta_2^2 U_{q_{e}} (U_{q_{e}} + U_{q_{h}}) + 
U_{q_{e}} (-q_{e} (q_{e} + q_{h}) U_{q_{e}} - 
e^{2 i\phi} q_{e} (q_{e} + q_{h}) U_{q_{h}} \\ 
&+ 
e^{i\phi} (\eta_1^2 - i \eta_1 (q_{e} - q_{h}) + 
q_{e} (q_{e} + q_{h})) (U_{q_{e}} + U_{q_{h}})) + \Delta (-1 + 
e^{i\phi}) \eta_2 (q_{h} U_{q_{e}} (1 + 
e^{i\phi} U_{q_{e}} U_{q_{h}})\\& - 
q_{e} ((-1 + e^{i\phi}) U_{q_{e}} + U_{q_{e}}^3 + 
e^{i\phi} U_{q_{h}} - (-1 + 
e^{i\phi}) U_{q_{e}}^2 U_{q_{h}}))) V_{q_{e}}\big\}/(\Lambda_{ss}  V_{q_{h}}),\\
t_ {q_{e}q_{e}2} &= -\big\{2 e^{i\phi/2} (-2 q_ {e} q_ {h} U_ {q_{e}}^2 + 
q_ {e}^2 U_ {q_{e}} U_ {q_{e}} + 
e^{i\phi} q_ {e}^2 U_ {q_{e}} U_ {q_{e}} +q_ {h}^2 U_ {q_{e}} U_ {q_{e}} + 
e^{i\phi} q_ {h}^2 U_ {q_{e}} U_ {q_{e}}  
-2 e^{i\phi} q_ {e} q_ {h} U_ {q_{e}}^2 \\&+
\Delta_ 0 (-1 + 
e^{i\phi}) \eta_ 2 (-q_ {e} (1 + U_ {q_{e}}^2) U_ {q_{e}} + 
q_ {h} U_ {q_{e}} (1 + U_ {q_{e}}^2)) + 
i \eta_ 1 (-(1 + e^{i\phi}) q_ {h} U_ {q_{e}} U_ {q_{e}} \\&+ 
q_ {e} (U_ {q_{e}}^2 + 
e^{i\phi} U_ {q_{e}}^2)))\big\}/ \Lambda_ {ss}, \\
t_ {q_{e}q_{h}2} &= -2  e^{i\phi/
	2} (U_ {q_{e}} ((-1 + e^{i\phi}) q_ {e} (q_ {e} - 
q_ {h}) (U_ {q_{e}} + U_ {q_{e}}) - 
I \eta_ 1 (q_ {e} U_ {q_{e}} + e^{i\phi} q_ {h} U_ {q_{e}} - 
e^{i\phi} q_ {e} U_ {q_{e}} \\
&-q_ {h} U_ {q_{e}})) + \Delta_ 0 (-1 + 
e^{i\phi}) \eta_ 2 (q_ {e} (-U_ {q_{e}}^3 + U_ {q_{e}}) + 
q_ {h} U_ {q_{e}} (-1 + 
U_ {q_{e}} U_ {q_{e}}))) V_ {q_{e}}/(\Lambda_ {ss} V_{q_{h}})\,,
\end{split}
\end{align}
Likewise, for the remaining three processes we obtain
\begin{align}
\label{eqsns3}
\begin{split}
r_ {q_{h}q_{e}1} &= -2 \big\{\Delta_ 0^2 (-1 + 
e^{i\phi})^2 \eta_ 2^2 U_ {q_{e}} (U_ {q_{e}} + U_ {q_{e}}) + 
U_ {q_{e}} (-e^{2 i\phi} q_ {h} (q_ {e} + q_ {h}) U_ {q_{e}} - 
q_ {h} (q_ {e} + q_ {h}) U_ {q_{e}}  \\
&+e^{i\phi} (\eta_ 1^2 - I \eta_ 1 (q_ {e} - q_ {h}) + 
q_ {h} (q_ {e} + q_ {h})) (U_ {q_{e}} + 
U_ {q_{e}})) \\
&+ \Delta_ 0 (-1 + 
e^{i\phi}) \eta_ 2 (U_ {q_{e}} (q_ {e} - 
q_ {h} (-1 + U_ {q_{e}} U_ {q_{e}} + U_ {q_{e}}^2)) + 
e^{i\phi} (q_ {e} U_ {q_{e}} U_ {q_{e}}^2  \\
&+q_ {h} (-U_ {q_{e}} + 
U_ {q_{e}} (-1 + 
U_ {q_{e}}^2))))\big\} V_ {q_{h}}/(\Lambda_ {ss} Vqe), \\
r_ {q_{h}q_{h}1} &= -\big\{(q_ {e}^2 - q_ {h}^2) U_ {q_{e}} U_ {q_{e}} - 
e^{2 i\phi} (q_ {e}^2 - 
q_ {h}^2) U_ {q_{e}} U_ {q_{e}} + \Delta_ 0^2 (-1 + 
e^{i\phi})^2 \eta_ 2^2 (U_ {q_{e}}^2 - U_ {q_{e}}^2)  \\
&+e^{i\phi} \eta_ 1 (\eta_ 1 - 2 i q_ {e}) (U_ {q_{e}}^2 - 
U_ {q_{e}}^2) + \Delta_ 0 (-1 + 
e^{i\phi}) \eta_ 2 (q_ {h} ((-1 + 
e^{i\phi}) U_ {q_{e}} + (-1 + 
e^{i\phi}) U_ {q_{e}}^2 U_ {q_{e}}  \\
&-(1 + 
e^{i\phi}) U_ {q_{e}} (-1 + U_ {q_{e}}^2)) +
q_ {e} (-(1 + e^{i\phi}) U_ {q_{e}} + (1 + 
e^{i\phi}) U_ {q_{e}}^2 U_ {q_{e}}  
-(-1 + 
e^{i\phi}) U_ {q_{e}} (1 + 
U_ {q_{e}}^2)))\big\}/(d \Lambda_ {ss}),\\
t_ {q_{h}q_{e}2} &= -\big\{2  e^{i\phi/
	2}  (U_ {q_{e}} (-(-1 + e^{i\phi}) (q_ {e} - 
q_ {h}) q_ {h} (U_ {q_{e}} + U_ {q_{e}}) - 
i \eta_ 1 (q_ {e} U_ {q_{e}} + e^{i\phi} q_ {h} U_ {q_{e}}  \\
&-e^{i\phi} q_ {e} U_ {q_{e}} - 
q_ {h} U_ {q_{e}})) + \Delta_ 0 (-1 + 
e^{i\phi}) \eta_ 2 (q_ {e} U_ {q_{e}} (-1 + 
U_ {q_{e}} U_ {q_{e}}) + q_ {h} (U_ {q_{e}} - 
U_ {q_{e}}^3))) V_ {q_{h}}\big\}/(\Lambda_ {ss} V_ {q_{e}}),\\
t_ {q_{h}q_{h}2} &= 2 e^{i\phi/
	2} \big\{e^{i\phi} U_ {q_{e}} (2 q_ {e} q_ {h} U_ {q_{e}} - 
q_ {e}^2 U_ {q_{e}} - q_ {h}^2 U_ {q_{e}} + 
i \eta_ 1 (q_ {h} U_ {q_{e}} - q_ {e} U_ {q_{e}})) + 
U_ {q_{e}} (-q_ {e}^2 U_ {q_{e}} - q_ {h}^2 U_ {q_{e}}  \\
&+2 q_ {e} q_ {h} U_ {q_{e}} - 
i \eta_ 1 (q_ {e} U_ {q_{e}} - 
q_ {h} U_ {q_{e}})) + \Delta_ 0 (-1 + 
e^{i\phi}) \eta_ 2 (-q_ {e} (1 + U_ {q_{e}}^2) U_ {q_{e}} + 
q_ {h} U_ {q_{e}} (1 + U_ {q_{e}}^2))\big\}/\Lambda_ {ss}\,,\\
r_ {q_{e}q_{e}2} &=  \big\{e^{2 i\phi} (q_ {e}^2 - 
q_ {h}^2) U_ {q_{e}} U_ {q_{e}} + (-q_ {e}^2 + 
q_ {h}^2) U_ {q_{e}} U_ {q_{e}} + \Delta_ 0^2 (-1 + 
e^{i\phi})^2 \eta_ 2^2 (U_ {q_{e}}^2 - U_ {q_{e}}^2)  \\
&+e^{i\phi} \eta_ 1 (\eta_ 1 + 2 i q_ {h}) (U_ {q_{e}}^2 - 
U_ {q_{e}}^2) - \Delta_ 0 (-1 + 
e^{i\phi}) \eta_ 2 (-q_ {h} ((-1 + 
e^{i\phi}) U_ {q_{e}} + (-1 + 
e^{i\phi}) U_ {q_{e}}^2 U_ {q_{e}} \\ 
&+(1 + 
e^{i\phi}) U_ {q_{e}} (-1 + U_ {q_{e}}^2)) + 
q_ {e} (-(1 + e^{i\phi}) U_ {q_{e}} + (1 + 
e^{i\phi}) U_ {q_{e}}^2 U_ {q_{e}} 
+ 
(e^{i\phi}-1 ) U_ {q_{e}} (1 + 
U_ {q_{e}}^2)))\big\}/\Lambda_ {ss},\\
r_ {q_{e}q_{h}2} &= -2 \big\{\Delta_ 0^2 (-1 + 
e^{i\phi})^2 \eta_ 2^2 U_ {q_{e}} (U_ {q_{e}} + U_ {q_{e}}) - 
U_ {q_{e}} (e^{2 i\phi} q_ {e} (q_ {e} + q_ {h}) U_ {q_{e}} 
+q_ {e} (q_ {e} + q_ {h}) U_ {q_{e}} \\&- 
e^{i\phi} (\eta_ 1^2 - i \eta_ 1 (q_ {e} - q_ {h}) + 
q_ {e} (q_ {e} + q_ {h})) (U_ {q_{e}} + 
U_ {q_{e}})) + \Delta_ 0 (e^{i\phi}-1) \eta_ 2 (-q_ {h} U_ {q_{e}} (e^{i\phi} + 
U_ {q_{e}} U_ {q_{e}}) \\&+ 
q_ {e} (U_ {q_{e}} - e^{i\phi} U_ {q_{e}} + 
e^{i\phi} U_ {q_{e}}^3 + 
U_ {q_{e}}  + (e^{i\phi}-1 ) U_ {q_{e}}^2 U_ {q_{e}}))\big\} V_ {q_{e}}/(\Lambda_
{ss} V_{q_{h}}),\\
t_ {q_{e}q_{e}1} &= 2 e^{i\phi/
	2}  \big\{U_ {q_{e}} (-q_ {e}^2 U_ {q_{e}} + 
i (\eta_ 1 + i q_ {h}) q_ {h} U_ {q_{e}} + 
q_ {e} (-i \eta_ 1 + 2 q_ {h}) U_ {q_{e}}) + 
e^{i\phi} U_ {q_{e}} (2 q_ {e} q_ {h} U_ {q_{e}}  
-q_ {e}^2 U_ {q_{e}} \\&- 
q_ {h}^2 U_ {q_{e}} - 
i \eta_ 1 (q_ {e} U_ {q_{e}} - 
q_ {h} U_ {q_{e}}))  
+ \Delta_ 0 (-1 + 
e^{i\phi}) \eta_ 2 (-q_ {e} (1 + U_ {q_{e}}^2) U_ {q_{e}} + 
q_ {h} U_ {q_{e}} (1 + U_ {q_{e}}^2))\big\}/\Lambda_ {ss}, \\
t_ {q_{e}q_{h}1} &= 2  e^{i\phi/
	2}\big\{ \Delta_ 0 (e^{i\phi}-1) \eta_ 2 (q_ {e} (U_ {q_{e}}-U_ {q_{e}}^3) + 
q_ {h} U_ {q_{e}} (U_ {q_{e}} U_ {q_{e}}-1)) + 
U_ {q_{e}} (i \eta_ 1 q_ {h} U_ {q_{e}}  
-q_ {e}^2 (U_ {q_{e}} + U_ {q_{e}}) \\&+ 
q_ {e} ( q_ {h} (U_ {q_{e}} + U_ {q_{e}})-i \eta_ 1 U_ {q_{e}}) + 
e^{i\phi} (q_ {e} (q_ {e} - q_ {h}) (U_ {q_{e}} + U_ {q_{e}}) + 
i \eta_ 1 (q_ {e} U_ {q_{e}}  
-q_ {h} U_ {q_{e}})))\big\} V_ {q_{e}}/(\Lambda_ {ss} V_ {q_{h}})\,,\\
r_ {q_{h}q_{e}2} &= -2\big\{ \Delta_ 0^2 (-1 + 
e^{i\phi})^2 \eta_ 2^2 U_ {q_{e}} (U_ {q_{e}} + U_ {q_{e}}) - 
U_ {q_{e}} (q_ {h} (q_ {e} + q_ {h}) U_ {q_{e}} + 
e^{2 i\phi} q_ {h} (q_ {e} + q_ {h}) U_ {q_{e}}  \\
&-e^{i\phi} (\eta_ 1^2 - i \eta_ 1 (q_ {e} - q_ {h}) + 
q_ {h} (q_ {e} + q_ {h})) (U_ {q_{e}} + 
U_ {q_{e}})) + \Delta_ 0 (-1 + 
e^{i\phi}) \eta_ 2 (-q_ {e} U_ {q_{e}} (e^{i\phi} + 
U_ {q_{e}} U_ {q_{e}})  \\
&+q_ {h} (U_ {q_{e}} - e^{i\phi} U_ {q_{e}} + 
e^{i\phi} U_ {q_{e}}^3 + 
U_ {q_{e}} (1 + (-1 + 
e^{i\phi}) U_ {q_{e}}^2))) V_ {q_{h}}\big\}/( \Lambda_
{ss} V_ {q_{e}}), \\
r_ {q_{h}q_{h}2} &= -\big\{e^{2 i\phi} (q_ {e}^2 - 
q_ {h}^2) U_ {q_{e}} U_ {q_{e}} + (-q_ {e}^2 + 
q_ {h}^2) U_ {q_{e}} U_ {q_{e}} + \Delta_ 0^2 (-1 + 
e^{i\phi})^2 \eta_ 2^2 (U_ {q_{e}}^2 - U_ {q_{e}}^2)  \\
&+e^{i\phi} \eta_ 1 (\eta_ 1 - 2 i q_ {e}) (U_ {q_{e}}^2 - 
U_ {q_{e}}^2) - \Delta_ 0 (-1 + 
e^{i\phi}) \eta_ 2 (-q_ {h} ((-1 + 
e^{i\phi}) U_ {q_{e}} + (-1 + 
e^{i\phi}) U_ {q_{e}}^2 U_ {q_{e}}  \\
&+ (1 + 
e^{i\phi}) U_ {q_{e}} (-1 + U_ {q_{e}}^2)) + 
q_ {e} (-(1 + e^{i\phi}) U_ {q_{e}}  + 
(1 + 
e^{i\phi}) U_ {q_{e}}^2 U_ {q_{e}}  
+(e^{i\phi}-1) U_ {q_{e}} (1 + 
U_ {q_{e}}^2)))\big\}/\Lambda_ {ss}, \\
t_ {q_{h}q_{e}1} &= 2 e^{i\phi} \big\{\Delta_ 0 (-1 + 
e^{i\phi}) \eta_ 2 (q_ {e} U_ {q_{e}} (-1 + U_ {q_{e}} U_ {q_{e}}) +
q_ {h} (U_ {q_{e}} - U_ {q_{e}}^3))  
+U_ {q_{e}} ((q_ {e} - q_ {h}) q_ {h} (U_ {q_{e}} + U_ {q_{e}}) \\&+ 
i \eta_ 1 (q_ {h} U_ {q_{e}} - q_ {e} U_ {q_{e}}) + 
e^{i\phi} (-(q_ {e} - q_ {h}) q_ {h} (U_ {q_{e}} + U_ {q_{e}}) 
+i \eta_ 1 (q_ {e} U_ {q_{e}} - 
q_ {h} U_ {q_{e}})))\big\} V_ {q_{h}}/(\Lambda_ {ss} V_ {q_{e}}), \\
t_ {q_{h}q_{h}1} &= -2 e^{i\phi/2} \big\{-2 q_ {e} q_ {h} U_ {q_{e}}^2 + 
q_ {e}^2 U_ {q_{e}} U_ {q_{e}} + 
e^{i\phi} q_ {e}^2 U_ {q_{e}} U_ {q_{e}} + 
q_ {h}^2 U_ {q_{e}} U_ {q_{e}} + 
e^{i\phi} q_ {h}^2 U_ {q_{e}} U_ {q_{e}}  
-2 e^{i\phi} q_ {e} q_ {h} U_ {q_{e}}^2 \\&+ \Delta_ 0 (-1 + 
e^{i\phi}) \eta_ 2 (-q_ {e} (1 + U_ {q_{e}}^2) U_ {q_{e}} + 
q_ {h} U_ {q_{e}} (1 + U_ {q_{e}}^2)) 
-i \eta_ 1 (-(1 + e^{i\phi}) q_ {e} U_ {q_{e}} U_ {q_{e}} + 
q_ {h} (U_ {q_{e}}^2 \\&+ 
e^{i\phi} U_ {q_{e}}^2))\big\}/ \Lambda_ {ss}\,,
\end{split}
\end{align}
where $\Lambda_{ss}$ in the denominator of the above expressions is given by
\begin{equation}\label{denoss}
\begin{split}
\Lambda_ {ss} &= (-(q_ {e} + q_ {h})^2 U_ {q_{e}} U_ {q_{e}} - 
e^{i\phi} (q_ {e} + 
q_ {h})^2 U_ {q_{e}} U_ {q_{e}} + \Delta_ 0^2 (-1 + 
e^{i\phi})^2 \eta_ 2^2 (U_ {q_{e}} + 
U_ {q_{e}})^2  \\
&+\Delta_ 0 (-1 + e^{i\phi})^2 \eta_ 2 (q_ {e} + 
q_ {h}) (-U_ {q_{e}} + U_ {q_{e}}^2 U_ {q_{e}} + 
U_ {q_{e}} (-1 + U_ {q_{e}}^2)) + 
e^{i\phi} (-2 q_ {e}^2 U_ {q_{e}} U_ {q_{e}} \\
&- 2 q_ {h}^2 U_ {q_{e}} U_ {q_{e}} + \eta_ 1^2 (U_ {q_{e}} + U_ {q_{e}})^2 - 
2 i \eta_ 1 (q_ {e} - q_ {h}) (U_ {q_{e}} + U_ {q_{e}})^2 + 
4 q_ {e} q_ {h} (U_ {q_{e}}^2 + U_ {q_{e}} U_ {q_{e}} + U_ {q_{e}}^2)))\,.
\end{split}
\end{equation}

The coefficients given by Eqs.\,(\ref{eqsns4}) and \,(\ref{eqsns3}) allow the full characterization of the scattering processes given by Eqs.\,(\ref{scatqelss}).

\subsection{Green's function in left S}
We are finally in the position to calculate the Green's function. Since we are primarily interested in the superconducting phase dependence of the pair amplitudes, it is enough to analyze either the Green's function on the left or right superconducting region. In the left S region, we obtain 
\begin{equation}
\label{SNSGAPP}
\begin{split}
G_{ee}^{r}(x,x',\omega)&= -s \Big[ U_{q_{e}} e^{i q_e |x-x'|} +  U_{q_{h}} e^{-i q_h |x-x'|}  \Big] +
s \Big[ U_{q_{e}} r_{q_{e}q_{e}1} e^{-i q_e (x+x')} + U_{q_{h}} r_{q_{h}q_{h}1} e^{i q_h (x+x')}  \Big] \\
&-s \Big[  \frac{V_{q_{h}} U_{q_{h}}}{V_{q_{e}}} r_{q_{e}q_{h}1} e^{i (q_h x - q_e x')}  + \frac{U_{q_{e}} V_{q_{e}} }{V_{q_{h}}} r_{q_{h}q_{e}1} e^{-i(q_e x - q_h x')} \Big] \,,\\
G_{eh}^{r}(x,x',\omega)&= s  \Big[ e^{-i|x-x'|q_h} - e^{i|x-x'|q_e} \Big]  {\rm sgn}(x-x') +
 s \Big[ -r_{q_{h}q_{h}1} e^{ i(x+x')q_h} +  r_{q_{e}q_{e}1}  e^{-i(x+x')q_e} \Big] \\
&+ s   \Big[ -\frac{V_{q_{h}}U_{q_{h}}}{V_{q_{e}}U_{q_{e}}} r_{q_{e}q_{h}1}  e^{i(q_hx - q_e x')} + \frac{V_{q_{e}}U_{q_{e}}}{V_{q_{h}}U_{q_{h}}} r_{q_{h}q_{e}1} e^{-i(q_ex - q_h x')} \Big]  \,,\\
G_{he}^{r}(x,x',\omega)&= s  \Big[ e^{-i|x-x'|q_h} - e^{i|x-x'|q_e} \Big]  {\rm sgn}(x-x') +
s \Big[ r_{q_{h}q_{h}1} e^{ i(x+x')q_h} -  r_{q_{e}q_{e}1}  e^{-i(x+x')q_e} \Big] \\
&+ s   \Big[ -\frac{V_{q_{h}}}{V_{q_{e}}} r_{q_{e}q_{h}1}  e^{i(q_hx - q_e x')} + \frac{V_{q_{e}}}{V_{q_{h}}} r_{q_{h}q_{e}1} e^{-i(q_ex - q_h x')} \Big]  \,,\\
G_{hh}^{r}(x,x',\omega)&= -s \Big[ \frac{1}{U_{q_{h}}} e^{-i q_h |x-x'|} +  \frac{1}{U_{q_{e}}} e^{i q_e |x-x'|}  \Big] -
s \Big[ \frac{1}{U_{q_{h}}} r_{q_{h}q_{h}1} e^{i q_h (x+x')} + \frac{1}{U_{q_{e}}}  r_{q_{e}q_{e}1} e^{-i q_e (x+x')}  \Big] \\
&-s \Big[  \frac{V_{q_{e}} }{V_{q_{h}} U_{q_{h}}} r_{q_{h}q_{e}1} e^{-i (q_e x - q_h x')}  + \frac{V_{q_{h}}  }{V_{q_{e}} U_{q_{e}}} r_{q_{e}q_{h}1} e^{i(q_h x - q_e x')} \Big]  \,,\\
\end{split}
\end{equation}
where $s=[i\eta (1+U_{q_{e}}^2)U_{q_{h}}]/[2 q_{h} (U_{q_{e}}^2 - U_{q_{h}}^2)]$, $\eta = 2m/\hbar^2$, and the scattering coefficients are given by Eqs.\,(\ref{eqsns4}-\ref{eqsns3}). Observe that the expressions above exhibit a component which is not space dependent at $x=x'$ and is thus a bulk quantity, while the other terms that are proportional to scattering coefficients imply their  interface nature. This is in agreement with the results for the S region in NS junction discussed before. 
Also note that the denominator of the scattering coefficients $\Lambda_{ss}$, given by Eq.\,(\ref{denoss}), is also denominator of all Green's functions elements (regular and anomalous). Hence, the poles of the Green's function, which correspond to the ABSs, coincide with the zeroes of $\Lambda_{ss}$ and  gives rise to the ABSs, as discussed in the main text.

\end{widetext}

\end{document}